\documentclass[11pt,a4paper]{article}
\pdfoutput=1
\usepackage{jheppub,amsmath,amssymb,color,youngtab,appendix,float}
\usepackage{graphicx}
\usepackage{booktabs}
\usepackage{comment}

\usepackage[utf8]{inputenc}
\usepackage[nottoc]{tocbibind}
\newcommand{\bea}{\begin{eqnarray}}
\newcommand{\eea}{\end{eqnarray}}
\newcommand{\Tr}{{\rm Tr}}

\usepackage{xcolor}

\usepackage{mathtools,stmaryrd,bm}
\makeatletter
\newcommand{\oset}[3][-0.25ex]{%
  \mathrel{\mathop{#3}\limits^{
    \vbox to#1{\kern-2\ex@
    \hbox{$\scriptstyle#2$}\vss}}}}
\makeatother

\DeclarePairedDelimiter{\sets}{\lbrace}{\rbrace}  
\DeclarePairedDelimiter{\pts}{(}{)}              
\DeclarePairedDelimiter{\ketv}{\lvert}{\rangle}   
\DeclarePairedDelimiter{\sq}{\lbrack}{\rbrack}   

\providecommand\given{}                          
\newcommand\SetSymbol[1][]{                      
	\nonscript\:#1\vert
	\allowbreak
	\nonscript\:
	\mathopen{}
}
\DeclarePairedDelimiterX\Sets[1]\{\}{
	\renewcommand\given{\SetSymbol[\delimsize]}
	#1
}
\DeclarePairedDelimiterX{\inner}[2]{\langle}{\rangle}{
	#1\,\delimsize\vert\,\mathopen{}#2
}
\DeclarePairedDelimiterX{\braketv}[3]{\langle}{\rangle}{
	#1\delimsize\vert\mathopen{}#2\delimsize\vert\mathopen{}#3
}

\newcommand{\lsum}[3][n]{\ensuremath{\sum\limits_{#1 = #2}^{#3}}}

\renewcommand\SetSymbol[1][]{                      
	\nonscript\:#1\vert
	\allowbreak
	\nonscript\:
	\mathopen{}
}
\DeclarePairedDelimiterX\Set[1]\{\}{
	\renewcommand\given{\SetSymbol[\delimsize]}
	#1
}
\makeatletter
\newcommand*{\transpose}{%
	{\mathpalette\@transpose{}}%
}
\newcommand*{\@transpose}[2]{%
	\raisebox{\depth}{$\m@th#1\intercal$}%
}
\makeatother
\usepackage{cleveref}
\begin{document}

\title{Finite-$N$ Operator Algebras and the Hilbert Space of Bilocal Holography}
\author[a,b]{Byoungjoon Ahn,}
\author[c,d]{Robert de Mello Koch,}\author[a]{Vatsal Garg,}
\author[e,f]{Antal Jevicki,}
\author[a,b,g]{ and Junggi Yoon,}
\affiliation[a]{Department of Physics, College of Science, Kyung Hee University, Seoul 02447, Republic of Korea}
\affiliation[b]{Research Institute for Basic Sciences, Kyung Hee University, Seoul 02447, Republic of Korea}
\affiliation[c]{School of Science, Huzhou University, Huzhou 313000, China}
\affiliation[d]{Mandelstam Institute for Theoretical Physics, School of Physics, University of the Witwatersrand, Private Bag 3, Wits 2050, South Africa}
\affiliation[e]{Department of Physics, Brown University,
182 Hope Street, Providence, RI 02912, United States}
\affiliation[f]{Brown Center for Theoretical Physics and Innovation, Brown University,
340 Brook Street, Providence, RI 02912, United States}
\affiliation[g]{International Research Center for Quantum Matter, Kyung Hee University, Seoul 02447, Republic of Korea}
\emailAdd{bjahn123@khu.ac.kr}
\emailAdd{robert@zjhu.edu.cn}
\emailAdd{vatsal\_garg@khu.ac.kr}
\emailAdd{antal\_jevicki@brown.edu}
\emailAdd{junggi.yoon@khu.ac.kr}
\date{January 2026}
\abstract{We give an operator-algebraic and representation-theoretic description of the Hilbert spaces of finite-\texorpdfstring{$N$}{N}  bilocal holography.  This work is a sequel to the finite-\texorpdfstring{$N$}{N} Hilbert space construction of \texorpdfstring{\href{https://arxiv.org/abs/2602.20788}{arXiv:2602.20788 [hep-th]}}{arXiv:2602.20788 [hep-th]}.  The central result is the establishment of an  invariant dual-pair operator algebra: before imposing the singlet constraint the Fock space carries commuting actions of the color group and of a bilocal Lie algebra, while the projection to the singlet sector selects a single irreducible representation of the invariant Lie algebra, which we call a \emph{master algebra}.  The finite-\texorpdfstring{$N$}{N} trace relations, beginning with the quadratic identities studied here, are shown to become representation-theoretic identities of the selected irreducible representation.  We summarize the orthogonal, symplectic and unitary cases, identify the corresponding finite-\texorpdfstring{$N$}{N} constraints, compute the singlet Casimirs, and explain how finite traces and partition functions are obtained through characters of the resulting irreducible representations. This provides a novel, previously unknown mathematical description of the singlet space.}

\maketitle

\section{Introduction}\label{sec:introduction}

Collective fields~\cite{Jevicki:1979mb} provide a constructive route from vector or matrix degrees of freedom to the invariant variables appropriate for holography~\cite{Maldacena:1997re,Gubser:1998bc,Witten:1998qj}.  In vector models the collective variables are bilocals, and the collective-field description is naturally organized as a large-$N$ expansion~\cite{deMelloKoch:1996mj,Das:2003vw,deMelloKoch:2010wdf,Das:2012dt,deMelloKoch:2018ivk}.  At finite $N$, however, the bilocal variables are overcomplete, but nevertheless the collective formalism remains valid~\cite{Das:2012dt}. A central role is played by finite-$N$ trace relations~\cite{Pr}, and these  relations reduce the naively infinite collective Hilbert space to the physical finite-$N$ singlet Hilbert space~\cite{deMelloKoch:2025ngs,deMelloKoch:2025rkw,deMelloKoch:2025eqt}.

The paper~\cite{deMelloKoch:2026dfo} studied this reduction from the point of view of finite-$N$ invariant theory, geometric quantization and explicit state counting.  The purpose of the present paper is to explain the same reduction from the operator-algebraic side.  The main result of our analysis is that the physical bilocal Hilbert space is a single irreducible representation of a large, constrained bilocal algebra, termed a \emph{master algebra}.  The trace relations are then the polynomial identities that hold inside this representation.  The finite-$N$ constraints, the Casimir eigenvalues, and the character formulae for the partition function are different aspects of one representation-theoretic structure.

The logic leading to this conclusion is completely transparent.  Before imposing the singlet constraint, the oscillator Fock space ${\cal F}$ carries two commuting actions.  One is the action of the color group $G$, and the other is generated by the invariant bilinears.  These actions form a reductive dual pair in the sense of Howe duality~\cite{Joung:2014qya,Joung:2015jza,Basile:2020qoe}.  The oscillator representation decomposes multiplicity freely, so the singlet sector ${\cal F}^G$ is paired with a unique irreducible representation of the bilocal algebra~\cite{Joung:2014qya,Joung:2015jza,Basile:2020qoe}.  This explains why the finite-$N$ bilocal Hilbert space is not an arbitrary direct sum of bilocal modules.

The paper is organized as follows.  In \cref{sec:algebra-summary} we summarize the bilocal operator algebras and fix the ordering conventions used in the body of the paper. We also interpret the finite-$N$ trace relations as Casimir identities.  In \cref{sec:irrep} we explain why the singlet Hilbert space is a single bilocal irreducible representation.  The fact that the Hilbert space is a single irreducible representation has an important consequence: traces over the finite-$N$ Hilbert space are naturally computed as characters of the corresponding bilocal algebra.  This gives a direct representation-theoretic interpretation of the partition function, as developed in \cref{sec:characters}. We conclude in \cref{sec:discussion} with a discussion of the implications of these results and several directions for future work. The appendices contain the explicit commutator calculations, collective-coordinate derivations and determinant identities.

\section{\texorpdfstring{Finite-$N$}{Finite-N} bilocal algebras: summary and conventions}\label{sec:algebra-summary}

In this section we give a complete summary of the bilocal operator algebras. 
The invariant bilocals are pair creation operators ($\bar A$), pair annihilation operators ($A$) and number operators ($B$ and $C$).  The symmetry of the pair operator determines the bilocal algebra.  For $O(N)$ bosons the color contraction is symmetric and the oscillators commute, so the pair operator is symmetric in the mode labels and the algebra is of symplectic type.  For $O(N)$ fermions the pair operator is antisymmetric and the algebra is of orthogonal type.  The compact symplectic color group reverses this pattern because the invariant tensor $\Omega_{ij}$ is antisymmetric.

Since we study a field theory, the theory has an infinite number of one-particle modes, which can be labelled, for example, by the momentum or by lattice site. To make the algebra and Hilbert space counting finite and well-defined, we regulate the number of one-particle modes to be $K$.  In what follows, mode labels are denoted by $k,l,m,n=1,\ldots,K$, and repeated color indices are summed.  Products such as $B^2$ and $\bar A A$ are matrix products over the mode indices; for example
\begin{equation}
        (B^2)_{kl}=\sum_{r=1}^K B_{kr}B_{rl},\qquad ( A\bar{A})_{kl}=\sum_{r=1}^K A_{kr}\bar{A}_{rl}\ .
\end{equation}
In the body of the paper we mainly use normal-ordered number operators.  Weyl-ordered formulae are collected in the appendices.  For example, in the $O(N)$ bosonic case
\begin{equation}
        B^{\rm W}_{kl}=B_{kl}+{\frac N2}\delta_{kl}, \qquad B_{kl}=\sum_{i=1}^N a_i^\dagger(k)a_i(l)\ ,
\end{equation}
and in the $O(N)$ fermionic case
\begin{equation}
        B^{\rm W}_{kl}=B_{kl}-{\frac N2}\delta_{kl}, \qquad B_{kl}=\sum_{i=1}^N b_i^\dagger(k)b_i(l)\ .
\end{equation}
This convention makes the finite-$N$ quadratic relations homogeneous. The pair annihilation and pair creation operators explicitly defined for the $O(N)$ bosonic case are,
\begin{equation}
    A_{kl} = \sum_{i=1}^N a_i(k)a_i(l),\qquad \bar A_{kl} = \sum_{i=1}^N a^\dagger_i(k)a^\dagger_i(l)\ ,
\end{equation}
and for the $O(N)$ fermionic case are,
\begin{equation}
    A_{kl} = \sum_{i=1}^N b_i(k)b_i(l),\qquad \bar A_{kl} = \sum_{i=1}^N b^\dagger_i(k)b^\dagger_i(l)\ .
\end{equation}
In the case of $O(N)$ bosons, $\bar{A}_{kl}$ is Hermitian conjugate of $A_{kl}$:
\begin{equation}
    (A_{kl})^\dagger = \bar A_{kl}\ ,
\end{equation}
but in the case of $O(N)$ fermions we see
\begin{equation}
    (A_{kl})^\dagger = -\bar A_{kl}\ .
\end{equation}
In general, we have
\begin{equation}
    (A_{kl})^\dagger = \varepsilon\bar A_{kl} \ .
\end{equation}
where the value $\varepsilon=+1$ gives the symplectic-type bilocal algebra, while $\varepsilon=-1$ gives the orthogonal-type bilocal algebra. The same parameter $\varepsilon$ also fixes the exchange symmetry of the pair operator in the mode labels,
\begin{equation}
 A_{kl}=\varepsilon A_{lk},\qquad \bar{A}_{kl}=\varepsilon \bar{A}_{lk}\ .
\end{equation}
Combining the conjugation rule with this symmetry, $A$ is the Hermitian conjugate of $\bar A$ as a matrix operator, \textit{i.e.}
\begin{equation}
    (A_{kl})^\dagger = \varepsilon\bar A_{kl}\,=\, \bar A_{lk} \ .
\end{equation}
The commutation relations are
\begin{gather}
[A_{kl},\bar A_{mn}]= B^{\rm W}_{\;\;mk}\delta_{nl}+ B^{\rm W}_{\;\;nl}\delta_{mk}+ \varepsilon B^{\rm W}_{\;\;ml}\delta_{nk}+ \varepsilon B^{\rm W}_{\;\;nk}\delta_{ml}\ ,\\
[A_{kl},A_{mn}]= 0 =  [\bar A_{kl},\bar A_{mn}]\ ,\\
[B^{\rm W}_{\;\;kl},B^{\rm W}_{\;\;mn}]=B^{\rm W}_{\;\;kn}\delta_{ml}-B^{\rm W}_{\;\;ml}\delta_{kn}\ ,\\
[A_{kl},B^{\rm W}_{\;\;mn}]=A_{kn}\delta_{lm}+\varepsilon A_{ln}\delta_{km}\ ,\\
[B^{\rm W}_{\;\;kl},\bar A_{mn}]=\bar A_{kn}\delta_{lm}+\varepsilon \bar A_{km}\delta_{ln}\ .
\end{gather}
The different microscopic theories correspond to different realizations of this universal algebra. For their central importance in specifying the finite-$N$ theory, we call the emergent Lie algebras: $\mathfrak{so}(2K)$, $\mathfrak{sp}(2K)$,\,\ldots by the name of \emph{master algebras} $G^*$. As advertised, it will be shown that it is one specific irreducible representation of the algebra $G^*$ that specifies the $O(N)$, $Sp(N)$ and $U(N)$ singlet Hilbert spaces of the theory.
Finally, it is interesting to note that in the passage to field theory we take $K\to\infty$. In this limit the emergent Lie algebras $G^*$ are $W_\infty$ algebras. 

\begin{table}[t]
\centering
\renewcommand{\arraystretch}{1.25}
\begin{tabular}{c c c c}
\toprule
Theory & symmetry of $\bar A_{kl}$ & bilocal algebra & finite-$N$ representation \\
\midrule
$O(N)$ bosons & symmetric & $\mathfrak{sp}(2K)$ & lowest-weight module \\
$O(N)$ fermions & antisymmetric & $\mathfrak{so}(2K)$ & highest-weight module \\
$Sp(N)$ bosons & antisymmetric & $\mathfrak{so}(2K)$ & orthogonal-type module \\
$Sp(N)$ fermions & symmetric & $\mathfrak{sp}(2K)$ & symplectic-type module \\
$U(N)$ bosons & rectangular & unitary type & rectangular module \\
$U(N)$ fermions & rectangular & unitary type & finite rectangular module \\
\bottomrule
\end{tabular}
\caption{The finite-$N$ bilocal algebras.  Statistics of the oscillators and symmetry of the color contraction determine whether the bilocal algebra is orthogonal or symplectic. The symmetry \emph{rectangular} refers to a highest-weight representation labelled by a rectangular Young diagram.}
\label{tab:bilocal-algebras}
\end{table}

\subsection{\texorpdfstring{Finite-$N$}{Finite-N}  relations}\label{subsec:finiteN-relations}

The master Lie algebra $G^*$ is compact in the fermionic case, and non-compact in the bosonic case. In both cases however, finite-$N$ constraints appear on the singlet Hilbert space in the form of quadratic relations.

In normal-ordered conventions, the basic finite-$N$ quadratic relations take the following compact form. For $O(N)$ bosons,
\begin{equation}
        \bar A A+(1+K-N)B-B^2=0\ . \label{eq:ON-boson-normal}
\end{equation}
For $O(N)$ fermions, 
\begin{equation}
        \bar A A-(N+K-1)B+B^2=0\ .  \label{eq:ON-fermion-normal}
\end{equation}
For $Sp(N)$ bosons, with $N$ even and with the compact symplectic invariant tensor $\Omega_{ij}$,
\begin{equation}
        \bar A A+(N-K+1)B+B^2=0\ .  \label{eq:SpN-boson-normal}
\end{equation}
For $Sp(N)$ fermions,
\begin{equation}
        \bar A A+(N+K+1)B-B^2=0\ . \label{eq:SpN-fermion-normal}
\end{equation}
For complex $U(N)$ oscillators one obtains two constraints because there are two number operators, $B$ and $C$.  In the bosonic case,
\begin{equation}
        \bar A A+(K-N)B-B^2=0\ ,\label{eq:UN-boson-normal1}
\end{equation}
\begin{equation}
        A\bar A-(C+N\mathbb I)(C+K\mathbb I)=0\ ,\label{eq:UN-boson-normal2}
\end{equation}
while in the fermionic case,
\begin{equation}
        \bar A A-(K+N)B+B^2=0\ ,\label{eq:UN-fermion-normal1}
\end{equation}
\begin{equation}
     A\bar A+(C-N\mathbb I)(C+K\mathbb I)=0 .\label{eq:UN-fermion-normal2}
\end{equation}
These quadratic, matrix-valued relations hold on the finite-$N$ singlet Hilbert space. Here is a simple derivation of the relations for the case of $O(N)$ bosons. The same argument extends straightforwardly to all other cases. At the quadratic level of operators, we make the ansatz that a linear combination of $\bar A A, B$ and $B^2$ vanish, to determine the form of the constraint. For the trivial vacuum state, it is easy to check that any linear combination satisfies the relation below
	\begin{equation}
		\pts*{c_1 \bar A_{kl} A_{lm} + c_2 B_{km} + c_3 B_{kl}B_{lm}}\ketv{0} = 0\ .
	\end{equation}
	For the first excited state, $\ketv{m_1n_1}\equiv \bar{A}_{m_1n_1}\ketv{0}$, we have
	\begin{gather}
		\pts*{c_1 \bar A_{kl} A_{lm} + c_2 B_{km} + c_3 B_{kl}B_{lm}}\bar A_{m_1n_1}\ketv{0} = 0\ ,\\
		\intertext{which implies that}
		Nc_1 + c_2 + c_3\pts*{1 + K} = 0\ .
	\end{gather}
	We repeat this for the next excited state, ${\ketv{m_1 m_2 n_1 n_2}\equiv \bar A_{m_1n_1}\bar A_{m_2n_2}\ketv{0}}$
	\begin{gather}
		\pts*{c_1 \bar A_{kl} A_{lm} + c_2 B_{km} + c_3 B_{kl}B_{lm}}\bar A_{m_1n_1}\bar A_{m_2n_2}\ketv{0} = 0\ ,\\
		\intertext{which holds only if}
		\begin{gathered}
			Nc_1 + c_2 + c_3\pts*{1 + K} = 0\ ,\\
		c_1 + c_3 = 0\,.
		\end{gathered}
	\end{gather}
	Notice that we have an additional condition for the second excited state, this allows us to fix the forms for $c_1, c_2$ and $c_3$ in terms of any one coefficient:
	\begin{equation}
		\begin{aligned}
			c_1 &= c_1\ ,\\
			c_2 &= \pts*{1 + K - N}c_1\ ,\\
			c_3 &= -c_1\ .
		\end{aligned}
	\end{equation}
	Finally, the constraint takes the following form:
	\begin{equation}
		\bar A A + \pts*{1 + K - N}B - B^2 = 0\ .\label{eq:ONBosonconstraint}
	\end{equation}
    More detailed commutator calculations leading to these formulae are given in appendix~\ref{app:operator-algebras}.  A rough sketch of a proof for \cref{eq:ONBosonconstraint} holding at all orders is given in appendix~\ref{subsec:ProofON}. The collective-coordinate derivations, including the Weyl-to-normal ordering conversion, are given in appendix~\ref{app:collective}.
The quadratic relations among the Lie algebra elements of
the master algebra $G^*$ play a central role in the reduction of the emergent Hilbert space. It should be noted that these are matrix valued relations. At the classical level (large $N$) they were observed in~\cite{Berezin:1978sn} in the coherent space representation of the invariant Hilbert space. At the quantum (finite $N$) level they were noted in~\cite{Jevicki:1980mj}. In what follows they will be used for complete specification of the finite-$N$ Hilbert space. This will come from a complete specification of all Casimirs of the algebra $G^*$.

\subsection{\texorpdfstring{Finite-$N$}{Finite-N} Casimirs}\label{subsec:finiteN-casimirs}

In the previous sections, we gave the quadratic relations obeyed by the bilocal algebras at finite $N$ and we argued that these relations select a specific finite-$N$ representation.  We now give further support for this conclusion by showing that the same quadratic trace relations can be interpreted as Casimir identities.  This provides a useful invariant characterization of the finite-$N$ representation. To spell out our specific conventions, we compute traces of powers of the matrix generators of the Lie algebra. These are known as Gelfand invariants and have been studied in detail in~\cite{PismaZhETF.1.15,PismaZhETF.2.34,1968IzMat}. In the $O(N)$ bosonic case, using Weyl-ordered generators, the quadratic combination
\begin{equation}
        C_2={\frac12}\Tr\left(A\bar A+\bar A A\right)-\Tr\left((B^{\rm W})^2\right)\ ,\label{eq:sp-casimir-def}
\end{equation}
has eigenvalue
\begin{equation}
        C_2={\frac{NK}2}\left(K+1-{\frac N2}\right)\ .\label{eq:sp-casimir-value}
\end{equation}
We can derive this from the finite-$N$ constraint that appears in \cref{eq:ON-boson-normal} and the symplectic algebra. Recall that
\begin{equation}
    \bar A_{kl}A_{lm} + (1 + K - N)B_{km} - B_{kl}B_{lm} = 0\ .\label{eq:explicit-matrix-ONBconstraint}
\end{equation}
From the above equations we can also obtain
\begin{equation}
    A_{kl}\bar A_{lm} - (1 + K + N)B_{mk} - B_{lk}B_{ml} = N(K+1)\delta_{kl}\ .\label{eq:explicit-matrix-ONBconstraintalt}
\end{equation}
Taking the trace of \cref{eq:explicit-matrix-ONBconstraint,eq:explicit-matrix-ONBconstraintalt} and adding gives,
\begin{equation}
    \frac12\Tr\left(\bar A A + A\bar A \right) - \Tr\left(B^2\right) - N\Tr(B) = \frac{NK}{2}(K+1)\ ,
\end{equation}
and
\begin{align}
    \frac12\Tr\left(\bar A A + A\bar A \right) - \Tr\left(B^2\right) - N\Tr(B) - \frac{N^2K}{4} &= \frac{NK}{2}(K+1) - \frac{N^2K}{4}\ ,\\[.5\baselineskip]
    \frac12\Tr\left(\bar A A + A\bar A \right) - \Tr\left(\left(B + \frac N2\mathbb I\right)^2\right) &= \frac{NK}{2}\left(K+1-\frac N2\right)\ .
\end{align}
This illustrates the concrete relation between the Casimir eigenvalue and the finite-$N$ constraint. We can easily extend this procedure to all other cases. 

For $O(N)$ fermions the relevant algebra is $\mathfrak{so}(2K)$.  The highest weight is $\lambda = \pts*{\frac N2}^K$ (with more details in \cref{subsec:on-fermions-hwm}).  With $\rho_i=K-i$, the quadratic Casimir is
\begin{equation}
        C_2^\Lambda=\sum_{i=1}^K \lambda_i(\lambda_i+2K-2i)\ . \label{eq:so-casimir-general}
\end{equation}
Substituting $\lambda_i=\frac N2$ gives
\begin{equation}
        C_2=\frac{NK}{2}\left(\frac N2+K-1\right)\ .  \label{eq:so-casimir-value}
\end{equation}
Higher Casimirs provide additional checks of the same representation assignment.  For example, the even Casimirs may be written as linear combinations of power sums~\cite{1968IzMat},
\begin{equation}
        S_{2m}^\Lambda=\sum_{i=1}^K\left[(\lambda_i+\rho_i)^{2m}-\rho_i^{2m}\right]\ .
        \label{eq:higher-casimir}
\end{equation}
These values are precisely reproduced using a collective field theory description. However the representation cannot be fully determined from these higher even Casimirs alone, the sign of the last component of the weight vector \emph{i.e.} in our case $\lambda_K$, cannot be fixed. To fix the representation uniquely we need to evaluate the Pfaffian Casimir,
\begin{equation}
    C'^\Lambda_K = (-1)^{\frac{K(K-1)}{2}}2^K K!\prod_{i=1}^K (\lambda_i + \rho_i)\ .
\end{equation}
It has the following property,
\begin{equation}
    C'_K(\lambda_1,\lambda_2,\ldots, \lambda_K) = -C'_K(\lambda_1,\lambda_2,\ldots, -\lambda_K)\ .
\end{equation}
Details on the explicit connection between higher Casimirs and Gelfand invariants can be found in \cref{subsec:on-fermions-hwm}.

For $U(N)$ fermions, we have a unitary algebra. The quadratic Casimir for the rectangular type representation $\lambda = N^K$ is
\begin{equation}
    C_2 = \frac{NK}{2}\left(\frac N2 + K\right)\ . 
\end{equation}
We have a general formula for the $n$\textsuperscript{th} Casimir,
\begin{equation}
    C_n = \frac{NK(N+2K)}{4(N+K)}\sets*{\pts*{\frac{N+2K}{2}}^{n-1} - \pts*{-\frac{N}{2}}^{n-1}}\ .
\end{equation}

\subsection{Summary}\label{subsec:conclusions}

As indicated an irreducible representation of the bilocal algebra is carried by the color-singlet Hilbert space.  The bilocal commutation relations determine the algebra of collective operators; the value of $N$ determines which orbit through this algebra is realized in the microscopic Fock space.  In this sense $N$ plays the role of a highest- or lowest-weight label: it fixes the extremal vector, the Casimir eigenvalues, and the null states that must be divided out.  The finite-$N$ trace relations are the equations saying that we have not obtained the universal enveloping representation freely generated by the bilocals, but only the particular representation that can be built from $N$ color degrees of freedom.  For fermions the restriction is transparent: after enough bilocal creation operators have acted, antisymmetrization forces the representation to terminate.  For bosons the representation does not terminate, but finite $N$ still fixes the representation by imposing nonlinear identities among the bilocals.  Thus the different finite-$N$ representations are different quantizations~\cite{Berezin:1975} of the same classical bilocal phase space, distinguished by the microscopic statistics and by the invariant tensor used to contract color indices.

\section{The singlet Hilbert space}\label{sec:irrep}

The most important structural result we establish is that the physical bilocal Hilbert space is a single irreducible representation of the bilocal algebra.  This point is conceptually important because it explains the finite-$N$ reduction in representation-theoretic terms: the reduction of the Hilbert space is built into the representation selected by the microscopic oscillator construction.  Finite $N$ prevents the bilocals from generating an unconstrained free algebra: certain combinations of bilocal operators vanish or become dependent, and these dependences are precisely the finite-$N$ trace relations. In the representation-theoretic description, the same relations appear as the statement that the physical states furnish a particular irreducible representation of the bilocal algebra, with $N$ fixing the extremal weight and the Casimir eigenvalues.

The mechanism behind this structure is the dual action~\cite{Basile:2020qoe} on the microscopic oscillator Fock space ${\cal F}$.  Before imposing the singlet constraint, ${\cal F}$ carries two mutually commuting symmetries: the color group $G$, acting on the color indices, and the bilocal algebra, generated by the invariant quadratic operators.  This is precisely the setting of Howe duality~\cite{Basile:2020qoe,Howe:1989nad}.  The resulting decomposition of ${\cal F}$ is multiplicity free: each irreducible representation of $G$ is paired with a unique irreducible representation of the bilocal algebra.  Consequently, when one projects onto the physical color-singlet sector ${\cal F}^G$, one does not obtain a sum of bilocal representations.  One obtains the single bilocal irreducible representation paired with the trivial representation of $G$. For a nice discussion of the relevant background, see~\cite{Basile:2020qoe}.

This statement will be made concrete through representative examples, where the abstract dual-pair argument is realized explicitly in the action of the bilocal generators on the singlet Hilbert space. The Fock vacuum is the extremal vector of the bilocal representation.  In the $O(N)$ bosonic case it is annihilated by the lowering operators and has Cartan eigenvalues $N/2$.  In the $O(N)$ fermionic case it is a highest-weight vector. The first example shows that the $O(N)$ fermion singlet sector is the highest-weight $\mathfrak{so}(2K)$ representation selected by the Fock vacuum.  The second shows that the $Sp(N)$ fermion bilocals realize the corresponding $\mathfrak{sp}(2K)$ algebra.  These examples make concrete the general dual-pair statement: after imposing the color-singlet constraint one obtains a single irreducible representation of the bilocal algebra.

\subsection{\texorpdfstring{$O(N)$}{O(N)} fermions and the
\texorpdfstring{$\mathfrak{so}(2K)$}{so(2K)}  module}\label{subsec:on-fermions-hwm}

Let $b_i(k)$ and $b_i^\dagger(k)$ obey
\begin{equation}
        \{b_i(k),b_j^\dagger(l)\}=\delta_{ij}\delta_{kl}\ .
\end{equation}
It is useful to introduce $2K$ real fermionic oscillators
\begin{equation}
        \chi_j^{2k-1}={\frac{1}{\sqrt 2}}\left(b_j(k)+b_j^\dagger(k)\right),\qquad
        \chi_j^{2k}={\frac{i}{\sqrt 2}}\left(b_j(k)-b_j^\dagger(k)\right)\ .
\end{equation}
They satisfy
\begin{equation}
        \{\chi_i^P,\chi_j^Q\}=\delta_{ij}\delta^{PQ}\ ,
\end{equation}
where $P, Q = 1,\ldots, 2K$. The color-singlet bilinears
\begin{equation}
        L_{PQ} = {\frac i2}\sum_{j=1}^{N}  [\chi_j^P,\chi_j^Q]\ , \label{eq:ONfermion-LAB}
\end{equation}
are antisymmetric in $P,Q$ and obey the $\mathfrak{so}(2K)$ algebra,
\begin{equation}
    [L_{PQ},L_{RS}] = iL_{PS}\delta_{QR} + iL_{QR}\delta_{PS} - iL_{PR}\delta_{QS} - iL_{QS}\delta_{PR}\ .
\end{equation}In terms of the original fermions, the Cartan generators may be chosen as
\begin{equation}
   H_k = L_{(2k-1)\,2k}={\frac N2}-\sum_{i=1}^{N}b_i^\dagger(k)b_i(k)\ .
        \label{eq:ONfermion-Cartan}
\end{equation}
It follows immediately that the Fock vacuum has weight
\begin{equation}
        H_k\left|0\right\rangle = {\frac N2}\left|0\right\rangle, \qquad \lambda_{\left|0\right\rangle} =
        \left({\frac N2},{\frac N2},\ldots,{\frac N2}\right)\ .  \label{eq:ONfermion-vacuum-weight}
\end{equation}
The relation to the bilocal generators used in previous sections is as follows.  Define
\begin{equation}
\begin{gathered}
   A_{kl}=\sum_{i=1}^{N}b_i(k)b_i(l),\qquad \bar A_{kl}=\sum_{i=1}^{N}b_i^\dagger(k)b_i^\dagger(l)\ ,\\
   B^{\rm W}_{\;\;kl}={\frac12}\sum_{i=1}^{N}\left( b_i^\dagger(k)b_i(l)-b_i(l)b_i^\dagger(k) \right)\ .
\end{gathered}
\end{equation}
The pair operators are antisymmetric,
\begin{equation}
        A_{kl}=-A_{lk}, \qquad \bar A_{kl}=-\bar A_{lk}\ ,
\end{equation}
as appropriate for an orthogonal bilocal algebra.  With the standard choice of positive roots of $\mathfrak{so}(2K)$, the positive-root generators are represented by $A$'s and by the appropriate off-diagonal $B^{\rm W}$'s.  They annihilate the vacuum
\begin{equation}
        A_{kl}\left|0\right\rangle=0, \qquad B^{\rm W}_{\;\;lk}\left|0\right\rangle=0 \quad (k<l)\ .
\end{equation}
Therefore $\left|0\right\rangle$ is a highest-weight vector of $\mathfrak{so}(2K)$ with highest weight
\begin{equation}
        \lambda= \left({\frac N2},{\frac N2},\ldots,{\frac N2}\right)\ .\label{eq:ONfermion-highest-weight}
\end{equation}
Gelfand invariants are defined as the traces of powers of the generators of the Lie algebra. They are proportional to the standard Casimir eigenvalues (up to some sign fixed by definition of the generator). For example, we have the second Gelfand invariant
\begin{equation}
    \Tr\left(L^2\right) = -\frac{NK}{2}(N + 2K - 2) = -2C_2\ .
\end{equation}
And we have the fourth Gelfand invariant
\begin{equation}
    \Tr\left(L^4\right) = \frac{NK}{8}(N + 2K - 2)(4(K-1)^2 + 2(K-1)N + N^2) = 2S_4 - 2(K-1)S_2 = 2C_4\ ,
\end{equation}
where $S$'s and $C$'s are the power sums and Casimir eigenvalues for the weight $\lambda = \pts*{\frac N2}^K$ respectively. Higher Gelfand invariants
are similarly related to higher order Casimirs. The Pfaffian Casimir is not evaluated in that manner, rather we have
\begin{equation}
    C'_K = \sum^K_{P_1,Q_1,\ldots,P_K,Q_K=1}\varepsilon_{\scriptscriptstyle P_1Q_1\cdots P_KQ_K}L_{P_1Q_1}L_{P_2Q_2}\cdots L_{P_KQ_K}\ .
\end{equation}
The physical $O(N)$ fermion singlet Hilbert space is the irreducible $\mathfrak{so}(2K)$ module generated from this highest-weight state. This gives a representation-theoretic interpretation of the finite-$N$ cutoff.  The bilocal creation operators $\bar A_{kl}$ which generate the representation become linearly dependent at finite $N$.  These linear dependences are precisely the finite-$N$ trace relations.

A simple check is to count the number of independent states generated by $\bar A$'s at fixed particle number.  The results for small values of $N$ and $K$ are shown in \cref{tab:Adim}.  These numbers agree with the
character computation.

\begin{table}[t]
\centering
\begin{tabular}{c|cccccccc}
\hline
\((N,K)\) & \((2,3)\) & \((2,4)\) & \((2,5)\) & \((2,6)\) &
\((3,2)\) & \((3,3)\) & \((3,4)\) & \((3,5)\)\\
\hline
\(n_P=0\)  & 1  & 1   & 1   & 1    & 1 & 1   & 1   & 1   \\
\(n_P=2\)  & 3  & 6   & 10  & 15   & 1 & 3   & 6   & 10  \\
\(n_P=4\)  & 6  & 21  & 55  & 120  & 1 & 6   & 21  & 55  \\
\(n_P=6\)  & 0  & 6   & 45  & 190  & 1 & 10  & 56  & 220 \\
\(n_P=8\)  &    & 1   & 15  & 120  &   & 0   & 21  & 225 \\
\(n_P=10\) &    &     & 0   & 15   &   &     & 6   & 126 \\
\(n_P=12\) &    &     &     & 1    &   &     & 1   & 35  \\
\(n_P=14\) &    &     &     &      &   &     &     & 0   \\
\hline
\end{tabular}
\caption{Number of the independent states generated by the bilocal creation
operators \(\bar A\) at fixed particle number \(n_P\).  Blank entries denote
levels which are not present for the corresponding values of \(N\) and \(K\).}
\label{tab:Adim}
\end{table}

At low levels, before trace relations appear, the $r=n_P/2$ bilocal pairs behave as commuting generators labelled by antisymmetric pairs of modes. The corresponding free count is
\begin{equation}
        d_{\rm free}(2r)=\binom{\binom{K}{2}+r-1}{r}\ . \label{eq:free-bilocal-count}
\end{equation}
Deviations from this formula at higher levels are due to the finite-$N$ relations. Thus the table gives a concrete state-counting manifestation of the fact that the singlet Hilbert space is a finite-dimensional highest-weight module, not the freely generated bilocal Fock space.
The raising operators annihilate $|0\rangle$, while repeated action with the allowed pair-creation operators generates the singlet Hilbert space.  Since the weight is finite, the fermionic module is finite-dimensional.  This is the algebraic origin of the finite Hilbert spaces counted in ref.~\cite{deMelloKoch:2026dfo}. 

\subsection{\texorpdfstring{$Sp(N)$}{Sp(N)} fermions and the
\texorpdfstring{$\mathfrak{sp}(2K)$}{sp(2K)} module}\label{subsec:rep-calc-spnfermion}

We now give the parallel check for $Sp(N)$ fermions.  Let $N$ be even, and let $\Omega_{ij}$ be the invariant antisymmetric tensor of $Sp(N)$,
\begin{equation}
        \Omega_{ij}=-\Omega_{ji}, \qquad \Omega^{ik}\Omega_{kj}=\delta^i{}_j\ .
\end{equation}
And our convention of raising and lowering of indices is as follows,
\begin{equation}
    v^i = \Omega^{ij}v_j,\quad v_i = \Omega_{ij}v^j\ .
\end{equation}
The fermionic oscillators now obey
\begin{equation}
        \{b_j(k),b^{\dagger j'}(l)\} = -i\delta_j{}^{j'}\delta_{kl}\ .
\end{equation}
The $Sp(N)$-invariant bilocals are defined as
\begin{equation}
\begin{gathered}
  A_{kl}=\Omega^{ij}b_i(k)b_j(l),\qquad \bar A_{kl}=\Omega_{ij}b^{\dagger i}(k)b^{\dagger j}(l)\ ,\\
  B^{\rm W}_{\;\;kl}={\frac i2}\sum_{j=1}^{N}\left( b^{\dagger j}(k)b_j(l)-b_j(l)b^{\dagger j}(k) \right)\ .
\end{gathered}
\end{equation}
Because the invariant tensor is antisymmetric and the oscillators are fermionic, the two minus signs cancel.  Hence
\begin{equation}
        A_{kl}=A_{lk}, \qquad \bar A_{kl}=\bar A_{lk}\ . \label{eq:SpNfermion-symmetric-pairs}
\end{equation}
This is the reason why the $Sp(N)$ fermion bilocal algebra is of symplectic type. To see this explicitly, introduce the real symplectic variables
\begin{equation}
        b_j(k) = {\frac 1{\sqrt 2}} \left(\chi^1_j(k)-i\chi^2_j(k)\right), \qquad
        b^\dagger_j(k) = {\frac 1{\sqrt 2}} \left(\chi^1_j(k)+i\chi^2_j(k)\right)\ .
\end{equation}
They obey
\begin{equation}
\{\chi^a_i(k),\chi^b_j(l)\}=\Omega_{ij}\,\varepsilon^{ab}\delta_{kl}\ , \label{eq:SpNfermion-chi-algebra}
\end{equation}
where $a,b = 1, 2$, and we fix $\varepsilon^{12}=1$. Now define
\begin{equation}
L_{ab}(k,l)={\frac i2}\Omega_{jj'}[\chi^{a\,j}(k),\chi^{b\,j'}(l)]\ .
\end{equation}
The generators satisfy
\begin{equation}
        L_{ab}(k,l)=L_{ba}(l,k)\ , \label{eq:SpNfermion-symmetry}
\end{equation}
and 
\begin{equation}
        (L_{ab}(k,l))^\dagger=L_{ba}(l,k)\ . \label{eq:SpNfermion-hermiticity}
\end{equation}
Now combine the pair $(a,k)$ into a single index $P=(a,k)$, with $P=1,\ldots,2K$. Let ${\cal J}_{PQ}$ be the $2K$-dimensional symplectic form
\begin{equation}
        {\cal J}_{(a,k)(b,l)}= \varepsilon^{ab}\delta_{kl}\ .
\end{equation}
Equivalently,
\begin{equation}\begin{gathered}
 {\cal J}_{kl}=0, \qquad {\cal J}_{k(K+l)}=\delta_{kl},\qquad {\cal J}_{(K+k)l}=-\delta_{kl}, \qquad
        {\cal J}_{(K+k)(K+l)}=0\ .
\end{gathered}\end{equation}
Writing $L_{PQ}$ for $L_{ab}(k,l)$, \cref{eq:SpNfermion-symmetry} becomes $L_{PQ}=L_{QP}$. Using \cref{eq:SpNfermion-chi-algebra}, one obtains
\begin{equation}
        [L_{PQ},L_{RS}] = -iL_{PS}{\cal J}_{QR}-iL_{QS}{\cal J}_{PR}-iL_{PR}{\cal J}_{QS}-iL_{QR}{\cal J}_{PS}\ .
        \label{eq:SpNfermion-sp2K}
\end{equation}
This is the standard oscillator realization of $\mathfrak{sp}(2K)$, up to the overall sign convention fixed by the definition of ${\cal J}$. There are two inequivalent choices for the Cartan subalgebra: $\sets[\Big]{L_{k(K+k)}}^K_{k=1}$ or  $\sets[\Big]{\frac12 \pts*{L_{kk}+L_{(K+k)(K+k)}}}^K_{k=1}$. We may pick the latter (denoted by $H_k$) and note that the weights of $\ketv{0}$ are
\begin{equation}
    \lambda = \pts*{-\frac{N}{2},-\frac{N}{2},\dots, -\frac{N}{2}}\ .
\end{equation}
One can also check as before that it is annihilated by all negative-root generators, and hence it is the lowest-weight vector of $\mathfrak{sp}(2K)$. Recall that $N$ must be even by definition, and hence this is an integer weight state. We can calculate the quadratic Gelfand invariant in this case too, but we must carefully consider the \emph{right} generator. For the symplectic Lie algebra, the generator is not actually symmetric\footnote{Recall \[\mathfrak{sp}(2K) = \Set*{X\in \mathfrak{gl}(2K)\given X^{\scriptscriptstyle\mathrm T}\! \mathcal J + \mathcal JX = 0}.\]
Check that if $Y \equiv \mathcal J X$, then $Y = Y^{\scriptscriptstyle\mathrm T}$.}. Rather the generator pre-multiplied by the symplectic form is symmetric. Therefore we can consider without loss of generality,
\begin{equation}
    \mathcal{L}_{PQ} \equiv -\mathcal{J}_{PR}L_{RQ}\ ,
\end{equation}
to be the generator of $\mathfrak{sp}(2K)$. We compute,
\begin{equation}
    \Tr\left(\mathcal{L}^2\right) = \Tr\left(\bar A A + A \bar A\right) - 2\Tr\left((B^{\rm W})^2\right) \propto C_2\ . 
\end{equation}
And the quadratic Casimir is, 
\begin{equation}
    C_2 = \frac12\Tr\left( \bar A A + A \bar A\right) - \Tr\left((B^{\rm W})^2\right) = -\frac{NK}{2}\left(K+1 + \frac{N}{2}\right)\ .
\end{equation}
Thus the $Sp(N)$ fermion bilocals realize the symplectic bilocal algebra.  The result is the direct analogue of the $O(N)$ boson case, with the important difference that the microscopic fermionic Fock space is finite-dimensional for fixed $N$ and $K$.  The calculation also explains why the $Sp(N)$ fermion case is paired with the $O(N)$ boson case in the classification of bilocal algebras: in both cases the invariant pair operators are symmetric in the mode labels.

\subsection{Summary}

The two examples in this section illustrate the representation-theoretic mechanism used in this work.  For $O(N)$ fermions, the color-singlet bilocals realize $G^* = \mathfrak{so}(2K)$, and the Fock vacuum is the highest-weight vector with weight
\begin{equation}
        \left({\frac N2},{\frac N2},\ldots,{\frac N2}\right)\ .
\end{equation}
For $Sp(N)$ fermions, the antisymmetric invariant tensor changes the symmetry of the pair operators, and the bilocals realize $G^* = \mathfrak{sp}(2K)$. For it  the Fock vacuum is the lowest-weight vector with weight
\begin{equation}
        \left(-{\frac N2},-{\frac N2},\ldots,-{\frac N2}\right)\ .
\end{equation}
In both cases, the singlet Hilbert space is spanned by a concrete irreducible representation of $G^*$.
Consequently, the finite-$N$ Hilbert space is not a freely generated bilocal Fock space.  It is the specific irreducible module selected by the microscopic oscillator construction and by the color-singlet constraint. We note, that in the continuum limit, when $K \to \infty$ we have $G^* = W_\infty$.

The same logic applies in the other cases.  What changes is the bilocal algebra and whether the representation is highest-weight, lowest-weight, or rectangular.  The trace relations in \cref{subsec:finiteN-relations} are precisely the polynomial identities satisfied by these representations.

\section{Traces, characters and partition functions}
\label{sec:characters}

We have argued that the finite-$N$ singlet Hilbert space is a single irreducible representation ${\cal R}_N$, of the
corresponding bilocal Lie algebra $G^*$.  This observation gives a particularly simple way to compute traces over the physical Hilbert space.  Consider the thermal partition function
\begin{equation}
        Z(\beta)=\Tr_{{\cal R}_N}\left(e^{-\beta H}\right)=\sum_{\alpha\in{\cal R}_N}e^{-\beta E_\alpha}\ ,
        \label{eq:thermal-partition-function}
\end{equation}
where $H|\alpha\rangle=E_\alpha|\alpha\rangle$. The representation-theoretic description becomes useful when the Hamiltonian
is chosen from the Cartan subalgebra of the bilocal algebra.  Let
$H_1,\ldots,H_r$ be a basis for the Cartan subalgebra, where $r$ is the rank
of the bilocal algebra.  We choose
\begin{equation}
        H=E_0+\sum_{p=1}^{r}h_p H_p\ , \label{eq:Cartan-Hamiltonian-general}
\end{equation}
where $E_0$ is a possible additive constant. A state of weight $\mu=(\mu_1,\ldots,\mu_r)$ obeys
\begin{equation}
        H_p|\mu,a\rangle = \mu_p|\mu,a\rangle, \qquad a=1,\ldots,m(\mu)\ , \label{eq:weight-state-definition}
\end{equation}
where $m(\mu)$ is the multiplicity of the weight $\mu$.  It follows that
\begin{align}
        Z(\beta) &= e^{-\beta E_0} \sum_{\mu}m(\mu) \exp\left(-\beta\sum_{p=1}^{r}h_p\mu_p\right), \nonumber\\
        &= e^{-\beta E_0} \chi_{{\cal R}_N}(x_1,\ldots,x_r)\ , \label{eq:partition-as-character}
\end{align}
where we have introduced the character of the representation ${\cal R}_N$
\begin{equation}
        \chi_{{\cal R}_N}(x_1,\ldots,x_r) = \Tr_{{\cal R}_N} \left( \prod_{p=1}^{r}x_p^{H_p} \right) =
        \sum_{\mu}m(\mu)\prod_{p=1}^{r}x_p^{\mu_p}\ . \label{eq:character-weight-expansion}
\end{equation}
This identity is the main point of the section.  A character is simply a trace of a group element in a specified representation.  Since the exponential of a Cartan element lies in the corresponding commuting subgroup -- more precisely, in its complexification for real positive $\beta$ -- the thermal trace becomes a character.  The finite-$N$ trace relations do not have to be imposed separately in evaluating \cref{eq:partition-as-character}: they are already encoded in the representation ${\cal R}_N$ and hence in its set of weights and their multiplicities.

It is often convenient to subtract the vacuum energy, to obtain a vacuum-normalized partition function $\widehat Z(\beta)$ as follows
\begin{equation}
        H_{\rm sub}=H-E_0,\qquad \widehat Z(\beta)=\Tr_{{\cal R}_N}\left(e^{-\beta H_{\rm sub}}\right)
        = e^{\beta E_0}Z(\beta)\ .\label{eq:vacuum-normalized-partition}
\end{equation}
With this convention the vacuum contributes the constant term $1$, and every remaining power of the Boltzmann variable counts an excitation above the vacuum.

For $U(N)$ fermions a Cartan Hamiltonian can be chosen as
\begin{equation}
        H_{U(N)} = \omega\sum_{p=1}^{K}\bigl(B_{pp}+C_{pp}\bigr)\ , \label{eq:UN-Cartan-Hamiltonian}
\end{equation}
where $B_{pq}\equiv b^{\dagger i}(p)b_i(q)$ and $C_{pq}\equiv d^\dagger_i(p)d^i(q)$ (See appendix~\ref{app:operator-algebras} for the details). The bilocal pair-creation operator then satisfies
\begin{equation}
        [H_{U(N)},\bar A_{pq}]=2\omega \bar A_{pq}\ .\label{eq:UN-pair-energy}
\end{equation}
Thus every action of $\bar A_{pq}$ raises the excitation energy by $2\omega$. The character of a group element $U\in U(N)$ in the irreducible representation labeled by Young diagram $R$ is given by the Schur polynomial $\chi_R(U)$. The singlet Hilbert space is in the rectangular representation, labeled by the Young diagram
\begin{equation}
\lambda=(N^K,0^K)=(\underbrace{N,\ldots,N}_{K},\underbrace{0,\ldots,0}_{K})\ ,
\label{eq:UN-rectangular-representation}
\end{equation}
of the unitary bilocal algebra acting on $2K$ variables.  Setting
\begin{equation}
    x=e^{-\beta\omega}, \qquad (x_1,\ldots,x_{2K})=(\underbrace{x,\ldots,x}_{K},\underbrace{x^{-1},\ldots,x^{-1}}_{K})\ ,
    \label{eq:UN-character-specialization}
\end{equation}
the relevant character is the Schur polynomial $s_{(N^K)}(x_1,\ldots,x_{2K})$.  The normalized partition function is
\begin{equation}
        \widehat Z^{\,f}_{U(N)}(x)=x^{KN}s_{(N^K)}\left(\underbrace{x,\ldots,x}_{K},\underbrace{x^{-1},\ldots,x^{-1}}_{K}
        \right)\ .\label{eq:UN-fermion-character-normalized}
\end{equation}
Although the Schur polynomial on the right-hand side is naturally a Laurent polynomial, the prefactor $x^{KN}$ shifts its lowest power to zero.  The result is therefore an ordinary polynomial in $x$, as required for a partition function whose vacuum has zero energy.

The same partition function was obtained in the matrix-integral formulation of the finite-$N$ Hilbert space in the form of the Hankel determinant
\begin{equation}
        D_{K,N}(x)=\det\left[\frac{1}{i+j+1}\,{}_2F_1\left( -N,i+j+1;i+j+2;1-x^2\right)\right]_{i,j=0}^{K-1}\ .
        \label{eq:UN-Hankel-determinant}
\end{equation}
The two apparently different expressions are related by
\begin{equation}
        D_{K,N}(x) = \frac{\displaystyle\prod_{m=1}^{K-1}(m!)^2}{\displaystyle\prod_{a,b=1}^{K}(N+a+b-1)}
        \widehat Z^{\,f}_{U(N)}(x)\ .\label{eq:UN-determinant-character-relation}
\end{equation}
Equivalently,
\begin{equation}
        \widehat Z^{\,f}_{U(N)}(x) = {\cal N}_{K,N}D_{K,N}(x), \qquad {\cal N}_{K,N}=
        \frac{\displaystyle\prod_{a,b=1}^{K}(N+a+b-1)}{\displaystyle\prod_{m=1}^{K-1}(m!)^2}\ .
        \label{eq:UN-determinant-normalization}
\end{equation}
The derivation of \cref{eq:UN-determinant-character-relation}, using the Weyl character formula and a confluent Vandermonde limit, is given in appendix~\ref{app:characters}.  Its conceptual meaning is simple: the matrix-integral determinant and the rectangular Schur polynomial are two different expressions for the same trace over the finite-$N$ singlet Hilbert space.

\begin{figure}[h]
\centering
\includegraphics[width=0.8\textwidth]{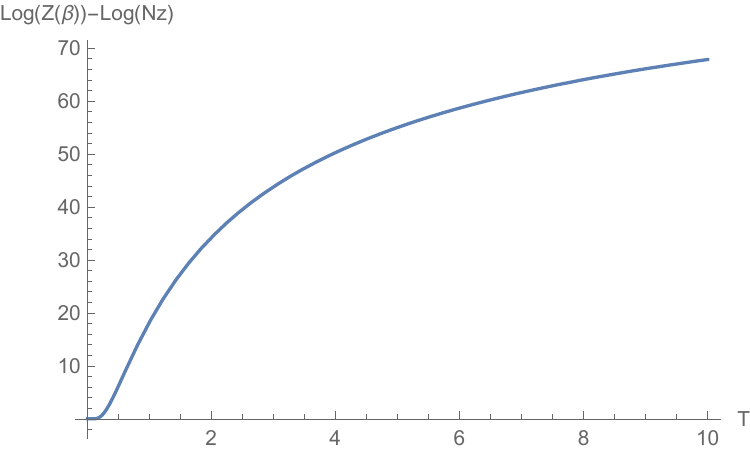}
\caption{A representative finite-$N$ partition-function plot.  The subtraction by the normalization factor isolates the temperature-dependent character contribution. $N_z={\cal N}_{K,N}$ is defined in \eqref{app:NKN-product}. The plot used the values $N=10$ and $K=12$.}
\label{fig:logZ}
\end{figure}

Figure~1 illustrates the thermal population of the finite-$N$ singlet Hilbert space. At low temperature the vacuum dominates and the partition function approaches one, so that $\log \widehat Z\to 0$. As the temperature is increased, more weights of the rectangular representation become thermally accessible and the partition function grows smoothly. At sufficiently high temperature all states in the finite-dimensional representation contribute with nearly equal weight, giving
\begin{equation}
\widehat Z(T\to\infty)=\dim{\cal R}_N,\qquad\log\widehat Z(T\to\infty)=\log\dim {\cal R}_N\ .
\end{equation}
The eventual saturation of the curve is therefore a thermodynamic manifestation of the finite-$N$ cutoff on the singlet Hilbert space. The smoothness of the curve is also expected at finite $N$; it represents a crossover as increasingly many states become populated, rather than a finite-temperature phase transition.

Finally, the equal-frequency for each mode specialization can be relaxed.  By retaining all Cartan generators independently, one may choose
\begin{equation}
        H =   \sum_{p=1}^{K}\omega_p H_p-E_0,  \qquad x_p=e^{-\beta\omega_p}\ ,
        \label{eq:unequal-frequency-Hamiltonian}
\end{equation}
and obtain the multivariable partition function
\begin{equation}
        Z(\beta;\omega_1,\ldots,\omega_K) = \chi_{{\cal R}_N}(x_1,\ldots,x_K)\ .
        \label{eq:unequal-frequency-character}
\end{equation}
Thus allowing the full set of Cartan fugacities gives a partition function in which the different one-particle modes may be assigned different frequencies.

The character formulae obtained in this section do more than provide an efficient method for evaluating partition functions.  They show that the finite-$N$ constraints have already been incorporated at the level of the representation itself.  Instead of summing over a freely generated bilocal Fock space and then imposing trace relations by hand, the character computes the trace directly over the irreducible bilocal representation selected by the microscopic singlet sector.  In this form the agreement between state counting, trace relations, Casimir identities, and thermodynamics becomes manifest.  The partition function is not just a generating function for degeneracies; it is a representation-theoretic invariant of the finite-$N$ Hilbert space.  This perspective should also be useful beyond the examples considered here, since different choices of statistics, color group, and one-particle spectrum can be incorporated by changing the corresponding bilocal representation and evaluating its character.

\section{Discussion}\label{sec:discussion}

The finite-$N$ reduction of the bilocal Hilbert space has two complementary descriptions.  The first, emphasized in ref.~\cite{deMelloKoch:2026dfo}, is the invariant-theory description: the collective variables obey trace relations, and implementing these relations reduces the Hilbert space.  The second, developed here, is the representation-theoretic description: the invariant bilocals generate a Lie algebra, and the singlet sector is the unique irreducible bilocal module selected by Howe duality.

This point of view clarifies the role of the trace relations.  They are not additional dynamical equations imposed on the collective theory.  They are identities internal to the finite-$N$ representation.  Their Casimir form fixes the representation, and their character form computes finite traces.  In this sense the finite-$N$ Hilbert space reduction, the quadratic constraints, the Casimir eigenvalues and the partition functions are all different expressions of the same algebraic fact.

The structure also explains why bosonic and fermionic vector models behave differently at finite $N$.  Fermionic singlet sectors are finite-dimensional because the selected highest-weight modules are finite.  Bosonic singlet sectors instead retain freely acting directions, with finite-$N$ trace relations cutting out the allowed representation-theoretic module.  The exchange between orthogonal and symplectic bilocal algebras in the $Sp(N)$ cases shows that this distinction is controlled jointly by statistics and by the symmetry of the color invariant tensor.

Once the partition function has been identified with the character of a finite-$N$ bilocal module, questions about entropy growth and thermodynamic limits become questions about the asymptotics of representations.  Large-$N$, large-$K$, and double-scaling limits can be studied by applying standard tools~\cite{Kerov} from character theory, Young-diagram asymptotics, and saddle-point analysis of group-theoretic formulae.  This gives a systematic framework for understanding how the finite-$N$ Hilbert space reduction affects the growth of states.  In particular, the emergence of entropy bounds, phase-transition-like behavior, or continuum collective-field limits can be phrased as asymptotic properties of the relevant bilocal characters.

There are several directions in which this work can be extended.  One is to make the connection with the Hironaka decomposition of the invariant ring more explicit at the level of the bilocal representation.  Another is to understand how these finite-$N$ algebraic identities appear in the bulk higher-spin description, especially in the organization of nonperturbative states and finite entropy.  The present paper supplies the algebraic starting point for these questions.

\begin{center} 
{\bf Acknowledgements}
\end{center}
We would like to acknowledge discussions with Euihun Joung. The work of A.J. is supported by the U.S. Department of Energy under contract DE-SC0010010. The work of RdMK was supported by a start up research fund of Huzhou University, a Zhejiang Province talent award and by a Changjiang Scholar award. J.Y. was supported by the National Research Foundation of Korea (NRF) grant funded by the Korean government (MSIT) (RS-2022-NR069038). J.Y. was supported by Brain Pool program funded by the Ministry of Science and ICT through the National Research Foundation of Korea (RS-2025-25457100). This work was supported by a grant from Kyung Hee University in 2025 (KHU-20251309).
B. A. was supported by Basic Science Research Program through the National Research Foundation of Korea (NRF) funded by the Korea government (MSIT) (RS-2026-25488427).

\appendix

\section{Operator algebras and finite-\texorpdfstring{$N$}{N} quadratic relations}
\label{app:operator-algebras}

In this appendix we collect the oscillator realizations of the bilocal algebras used in the main text and provide a proof for \cref{eq:ONBosonconstraint} holding at all orders in the singlet Hilbert space.  The one-particle Hilbert space is regulated to have dimension $K$, so all mode labels take values $k,l,m,n=1,\ldots,K$. There are two closely related conventions for the number operator.  In the real $O(N)$ and $Sp(N)$ cases it is convenient to use the Weyl-ordered bilocal $B^{\rm W}$. The Weyl-ordered form makes the Lie algebra covariance transparent, while the normal-ordered form gives homogeneous finite-$N$ trace relations. For generators constructed using only creation or only annihilation operators, normal-ordered and Weyl-ordered expressions agree, so the superscript is unnecessary. 

\subsection{Real vector models: orthogonal and symplectic bilocal algebras}

For real vector models the basic bilocals are pair annihilation, pair creation,
and number operators.  The sign structure of the algebra is controlled by the
symmetry of the pair operator in the mode labels.  We write
\begin{equation}
 A_{kl}=\varepsilon A_{lk},\qquad \bar A_{kl}=\varepsilon \bar A_{lk}, \qquad \text{with\;}\varepsilon=\pm 1\ .
\end{equation}
The value $\varepsilon=+1$ gives the symplectic-type bilocal algebra, while $\varepsilon=-1$ gives the orthogonal-type bilocal algebra. The commutation relations are
\begin{gather}
[A_{kl},\bar A_{mn}]= B^{\rm W}_{\;\;mk}\delta_{nl}+ B^{\rm W}_{\;\;nl}\delta_{mk}+\varepsilon B^{\rm W}_{\;\;ml}\delta_{nk}+\varepsilon B^{\rm W}_{\;\;nk}\delta_{ml}\ ,\label{app:real-universal-algebra1}\\
[A_{kl},A_{mn}]= 0 =  [\bar A_{kl},\bar A_{mn}]\ ,\label{app:real-universal-algebra2}\\
[B^{\rm W}_{\;\;kl},B^{\rm W}_{\;\;mn}]=B^{\rm W}_{\;\;kn}\delta_{ml}-B^{\rm W}_{\;\;ml}\delta_{kn}\ ,\label{app:real-universal-algebra3}\\
[A_{kl},B^{\rm W}_{\;\;mn}]=A_{kn}\delta_{lm}+\varepsilon A_{ln}\delta_{km}\ ,\label{app:real-universal-algebra4}\\
[B^{\rm W}_{\;\;kl},\bar A_{mn}]=\bar A_{kn}\delta_{lm}+\varepsilon \bar A_{km}\delta_{ln}\ . \label{app:real-universal-algebra5}
\end{gather}
The different microscopic theories correspond to different realizations of this universal algebra. It should be noted that in the above expressions we used $\bar A$ for the orthogonal-type bilocal algebra too, even though we use $\bar A$ in the following sections with orthogonal algebras.

\subsubsection{\texorpdfstring{$O(N)$}{O(N)} bosons}

The bosonic oscillators obey
\begin{equation}
        [a_i(k),a^\dagger_j(l)]=\delta_{ij}\delta_{kl}\ .
\end{equation}
The $O(N)$-invariant bilocals are
\begin{equation}
\begin{gathered}
        A_{kl}=\sum_{i=1}^{N}a_i(k)a_i(l),\qquad
        \bar A_{kl}=\sum_{i=1}^{N}a_i^\dagger(k)a_i^\dagger(l)\ ,\\
        B^{\rm W}_{\;\;kl}={\frac 12}\sum_{i=1}^{N}\left(a_i^\dagger(k)a_i(l)+a_i(l)a_i^\dagger(k)\right)\ .
\end{gathered}
\end{equation}
Here $A_{kl}=A_{lk}$, so $\varepsilon=+1$.  The bilocal algebra is therefore of symplectic type.
The Weyl-ordered finite-$N$ identity is
\begin{equation}
   \bar A A+(1+K)B^{\rm W}-(B^{\rm W})^2={\frac N2}\left(K+1-{\frac N2}\right){\mathbb I}\ .
   \label{app:ON-boson-weyl}
\end{equation}
Using $B^{\rm W}=B+{\frac N2}{\mathbb I}$, this becomes the homogeneous normal-ordered relation
\begin{equation}
        \bar A A+(1+K-N) B-B^2=0\ . \label{app:ON-boson-normal}
\end{equation}

\subsubsection{Proof for the constraint}\label{subsec:ProofON}
It is natural to ask whether \cref{eq:ONBosonconstraint} holds for arbitrary order excited states. We sketch a proof showing that indeed, it holds true at all orders. For this proof, let capital indices denote the modes \emph{i.e.} $I,J,K,L,M,N,\ldots = 1,\ldots, K$. We also define a useful object for this proof, $\sq*{\mathcal{A}, \mathcal{B}_1\mathcal{B}_2\cdots\mathcal{B}_n}_i \equiv \mathcal{B}_1\cdots\mathcal{B}_{i-1}\sq*{\mathcal{A}, \mathcal{B}_i}\mathcal{B}_{i+1}\cdots\mathcal{B}_{n}$. This implies:
	\begin{equation}
		\sq*{\mathcal{A}, \mathcal{B}_1\mathcal{B}_2\cdots\mathcal{B}_n} = \lsum[i]{1}{n} \sq*{\mathcal{A}, \mathcal{B}_1\mathcal{B}_2\cdots\mathcal{B}_n}_i\ .\label{eq:Leibniz}
	\end{equation}
	Consider the $n\textsuperscript{th}$ excited state, $\bar A_{M_1N_1}\bar A_{M_2N_2}\cdots \bar A_{M_nN_n}\ketv{0}$. We have the expression:
	\begin{equation}
		\pts*{c_1 \bar A_{IK}A_{KJ} + c_2 B_{IJ} + c_3 B_{IK}B_{KJ}}\bar A_{M_1N_1}\bar A_{M_2N_2}\cdots \bar A_{M_nN_n}\ketv{0}\ ,
	\end{equation}
	which is equivalent to,
	\begin{equation}
	\begin{split}
		\bar A_{M_1N_1}\bar A_{M_2N_2}\cdots \bar A_{M_nN_n}\pts*{c_1 \bar A_{IK}A_{KJ} + c_2 B_{IJ} + c_3 B_{IK}B_{KJ}}\ketv{0}~+\\ \pts*{c_1\sq*{\bar A_{IK}A_{KJ}, \bar A\cdots \bar A} + c_2\sq*{B_{IJ}, \bar A\cdots \bar A} + c_3\sq*{B_{IK}B_{KJ}, \bar A\cdots \bar A}}\ketv{0} \ ,
	\end{split}
	\end{equation}
	it is easy to see that the first term vanishes and all that remains are the commutators. We now apply the Leibniz rule (\cref{eq:Leibniz}) to each commutator and compare the terms. We have,
	\begin{align}
		&c_1\sq*{\bar A_{IK}A_{KJ}, \bar A_{M_1N_1}\cdots \bar A_{M_nN_n}}_k\notag\\
		&= \bar A_{M_1N_1}\cdots \bar A_{M_{k-1}N_{k-1}}\sq*{\bar A_{IK}A_{KJ}, \bar A_{M_kN_k}}\bar A_{M_{k+1}N_{k+1}}\cdots \bar A_{M_{n}N_{n}}\ ,\\
		\begin{split}
			&= \bar A\cdots \bar A\left(N\pts*{\bar A_{IN_k}\delta_{JM_k} + \bar A_{IM_k}\delta_{JN_k}} +\right.\\
			&\quad\left.+\pts*{\bar A_{IK}\delta_{M_kJ}B_{N_kK} + \bar A_{IK}\delta_{N_kJ}B_{M_kK} + \bar A_{IN_k}B_{M_kJ} + \bar A_{IM_k}B_{N_kJ}}\right)\bar A\cdots \bar A\ ,
		\end{split}
	\end{align}
	\begin{align}
		&c_2\sq*{B_{IJ}, \bar A_{M_1N_1}\cdots \bar A_{M_nN_n}}_k \notag\\
		&=\bar A_{M_1N_1}\cdots \bar A_{M_{k-1}N_{k-1}}\pts*{\bar A_{IN_k}\delta_{JM_k} + \bar A_{IM_k}\delta_{JN_k}}\bar A_{M_{k+1}N_{k+1}}\cdots \bar A_{M_{n}N_{n}}\ ,		
	\end{align}
	\begin{align}
		&c_3\sq*{B_{IJ}B_{JK}, \bar A_{M_1N_1}\cdots \bar A_{M_nN_n}}_k\notag\\
		\begin{split}
			&= \bar A\cdots \bar A\left(\pts*{\bar A_{IN_k}\delta_{JM_k} + \bar A_{IM_k}\delta_{JN_k}}\pts*{1 + \delta_{KK}} + \bar A_{IM_k}B_{N_kJ} + \bar A_{IN_k}B_{M_kJ}\right.\\
			&\quad\left.+~\bar A_{KN_k}\delta_{M_kJ}B_{IK} + \bar A_{KM_k}\delta_{N_kJ}B_{IK}\right)\bar A\cdots \bar A\ .
		\end{split}
	\end{align}
	We see that almost everything vanishes if \cref{eq:ONBosonconstraint} holds. The only remainder (at $k\textsuperscript{th}$ index) is:
	\begin{equation}
		\begin{split}
			\sq*{\cdots, \bar A\cdots \bar A}_k = c_1 \bar A_{M_1N_1}\cdots \bar A_{M_{k-1}N_{k-1}}\bar A_{IK}\pts*{\delta_{M_k J}B_{N_kK} + \delta_{N_kJ}B_{M_kK}}\bar A_{M_{k+1}N_{k+1}}\cdots \bar A_{M_nN_n}\\[.5\baselineskip]
			+~c_3 \bar A_{M_1N_1}\cdots \bar A_{M_{k-1}N_{k-1}}\pts*{\bar A_{KN_k}\delta_{M_kJ} + \bar A_{KM_k}\delta_{N_kJ}}B_{IK}\bar A_{M_{k+1}N_{k+1}}\cdots \bar A_{M_{n}N_{n}}\ .
		\end{split}
	\end{equation}
	The first term can be written as:
	\begin{align}
		\begin{split}
			\sq{c_1}_k &= c_1\bar A_{IJ}\bar A_{M_1N_1}\cdots \bar A_{M_nN_n}\pts*{\delta_{M_k J}B_{N_kK} + \delta_{N_kJ}B_{M_kK}} +\\ 
			&+ c_1\bar A_{M_1N_1}\cdots \bar A_{M_{k-1}N_{k-1}} \lsum[l]{1}{n-k} \bar A_{IK}\sq*{\delta_{M_k J}B_{N_kK} + \delta_{N_kJ}B_{M_kK}, \bar A_{M_{k+1}N_{k+1}}\cdots \bar A_{M_nN_n}}_l\ .
		\end{split}
	\end{align}
	As usual, the first line on the RHS above vanishes because the number operator is shifted to the rightmost end. Let the summand be denoted by $\sq*{c_1}_{kl}$ for brevity. For the second term, we similarly have:
	\begin{align}
		\begin{split}
			\sq{c_3}_k &= c_3 \bar A_{M_1N_1}\cdots \bar A_{M_nN_n}\pts*{\bar A_{KN_k}\delta_{M_k J} + \bar A_{KM_k}\delta_{N_kJ}}B_{IK} +\\ 
			&+ c_3 \bar A_{M_1N_1}\cdots \bar A_{M_{k-1}N_{k-1}} \lsum[l]{1}{n-k} \pts*{\bar A_{KN_k}\delta_{M_kJ} + \bar A_{KM_k}\delta_{N_kJ}}\sq*{B_{IK}, \bar A_{M_{k+1}N_{k+1}}\cdots \bar A_{M_nN_n}}_l\ .
		\end{split}
	\end{align}
	The first line on the RHS vanishes, let us similarly denote the summand as $\sq*{c_3}_{kl}$ for brevity. 
	\begin{align}
		\begin{split}
			\sq*{c_1}_{kl} &= \bar A_{M_{k+1}N_{k+1}}\cdots \bar A_{M_{l-1}N_{l-1}}\left[\bar A_{IM_l}\pts*{\bar A_{N_kN_l}\delta_{M_kJ} + \bar A_{M_kN_l}\delta_{N_kJ}} + \right.\\
			&\left. \qquad+\bar A_{IN_l}\pts*{\bar A_{N_kM_l}\delta_{M_kJ} + \bar A_{M_kM_l}\delta_{N_kJ}} \right]\bar A_{M_{l+1}N_{l+1}}\cdots \bar A_{M_{n}N_{n}}\ .
		\end{split}
	\end{align}
	and 
	\begin{align}
		\begin{split}
			\sq*{c_3}_{kl} &= \bar A_{M_{k+1}N_{k+1}}\cdots \bar A_{M_{l-1}N_{l-1}}\left[\bar A_{IM_l}\pts*{\bar A_{N_lN_k}\delta_{M_kJ} + \bar A_{N_lM_k}\delta_{N_kJ}} + \right.\\
			&\left. \qquad+\bar A_{IN_l}\pts*{\bar A_{M_lN_k}\delta_{M_kJ} + \bar A_{M_lM_k}\delta_{N_kJ}} \right]\bar A_{M_{l+1}N_{l+1}}\cdots \bar A_{M_{n}N_{n}}\ .
		\end{split}
	\end{align}
	Clearly both terms are identical, especially considering that $\bar A$ is symmetric with respect to mode indices. Therefore even the remainder vanishes if \cref{eq:ONBosonconstraint} holds.

\subsubsection{\texorpdfstring{$O(N)$}{O(N)} fermions}

The fermionic oscillators obey
\begin{equation}
        \{b_i(k),b^\dagger_j(l)\}=\delta_{ij}\delta_{kl}\ .
\end{equation}
The $O(N)$-invariant bilocals are
\begin{equation}
\begin{gathered}
  A_{kl}=\sum_{i=1}^{N}b_i(k)b_i(l),\qquad \bar A_{kl}=\sum_{i=1}^{N}b_i^\dagger(k)b_i^\dagger(l)\ ,\\
  B^{\rm W}_{\;\;kl}={\frac 12}\sum_{i=1}^{N}\left(b_i^\dagger(k)b_i(l)-b_i(l)b_i^\dagger(k)\right)\ .
\end{gathered}
\end{equation}
Now $A_{kl}=-A_{lk}$, so $\varepsilon=-1$.  The bilocal algebra is therefore of orthogonal type. The Weyl-ordered finite-$N$ identity is
\begin{equation}
  \bar A A+(1-K)B^{\rm W}+(B^{\rm W})^2={\frac N2}\left(K-1+{\frac N2}\right){\mathbb I}\ .\label{app:ON-fermion-weyl}
\end{equation}
In terms of normal-ordered operators, this is equivalently
\begin{equation}
        \bar A A-(N+K-1)B+ B^2=0\ .\label{app:ON-fermion-normal}
\end{equation}

\subsubsection{\texorpdfstring{$Sp(N)$}{Sp(N)} bosons}

For $Sp(N)$, with $N$ even, let $\Omega_{ij}$ be the invariant antisymmetric
tensor,
\begin{equation}
        \Omega_{ij}=-\Omega_{ji},  \qquad  \Omega^{ik}\Omega_{kj}=\delta^i{}_j\ .
\end{equation}
Indices are raised and lowered with $\Omega$ with the convention fixed in the main text.
The bosonic oscillators obey
\begin{equation}
        [a_j(k),a^{\dagger j'}(l)]= -i\delta_j{}^{j'}\delta_{kl}\ .
\end{equation}
The singlet bilocals are
\begin{equation}
\begin{gathered}
        A_{kl}=\Omega^{ij}a_i(k)a_j(l),  \qquad \bar A_{kl}=\Omega_{ij}a^{\dagger i}(k)a^{\dagger j}(l)\ ,\\
        B^{\rm W}_{\;\;kl}={\frac i2}\sum_{j=1}^{N}\left(a^{\dagger j}(k)a_j(l)+a_j(l)a^{\dagger j}(k)\right)\ .
\end{gathered}
\end{equation}
Because the oscillators commute but $\Omega_{ij}$ is antisymmetric we see,
\begin{equation}
        A_{kl}=-A_{lk},  \qquad   \bar A_{kl}=-\bar A_{lk}\ .
\end{equation}
Thus $\varepsilon=-1$, and the bilocal algebra is of orthogonal type. The Weyl-ordered finite-$N$ identity is
\begin{equation}
        \bar A A+(1-K)B^{\rm W}+(B^{\rm W})^2=-{\frac N2}\left(K-1-{\frac N2}\right){\mathbb I}\ .
        \label{app:SpN-boson-weyl}
\end{equation}
Equivalently, with normal-ordered operators,
\begin{equation}
        \bar A A+(N-K+1) B + B^2=0\ . \label{app:SpN-boson-normal}
\end{equation}

\subsubsection{\texorpdfstring{$Sp(N)$}{Sp(N)} fermions}

The fermionic oscillators obey
\begin{equation}
        \{b_j(k),b^{\dagger j'}(l)\}= -i\delta_j{}^{j'}\delta_{kl}\ .
\end{equation}
The $Sp(N)$-invariant bilocals are
\begin{equation}
\begin{gathered}
 A_{kl}=\Omega^{ij}b_i(k)b_j(l), \qquad \bar A_{kl}=\Omega_{ij}b^{\dagger i}(k)b^{\dagger j}(l)\ ,\\
 B^{\rm W}_{\;\;kl}={\frac i2}\sum_{j=1}^{N}\left(b^{\dagger j}(k)b_j(l)-b_j(l)b^{\dagger j}(k)\right)\ .
\end{gathered}
\end{equation}
Here the antisymmetry of $\Omega_{ij}$ is compensated by the fermion exchange sign, so
\begin{equation}
        A_{kl}=A_{lk}, \qquad \bar A_{kl}=\bar A_{lk}\ .
\end{equation}
Thus $\varepsilon=+1$, and the bilocal algebra is of symplectic type. The Weyl-ordered finite-$N$ identity is
\begin{equation}
        \bar A A+(1+K)B^{\rm W}-(B^{\rm W})^2= -{\frac N2}\left(K+1+{\frac N2}\right){\mathbb I}\ .
        \label{app:SpN-fermion-weyl}
\end{equation}
Equivalently, with normal-ordered operators,
\begin{equation}
        \bar A A+(N+K+1) B- B^2=0\ . \label{app:SpN-fermion-normal}
\end{equation}

\subsection{Complex vector models: unitary-type bilocal algebras}

For $U(N)$ models one has fundamental and antifundamental oscillators.  The bilocal algebra is generated by $A,\bar A$ together with two number-operator matrices $B$ and $C$.  The bosonic and fermionic cases have the same structural form, but differ by signs in the mixed commutator and in the finite-$N$ quadratic relations.

\subsubsection{\texorpdfstring{$U(N)$}{U(N)} bosons}

The bosonic oscillators obey
\begin{equation}
  [a_i(k),a^{\dagger j}(l)]=\delta_i{}^j\delta_{kl},\qquad  [b^i(k),b^\dagger_j(l)]=\delta^i{}_j\delta_{kl}\ ,
\end{equation}
with all other commutators vanishing.  The $U(N)$-invariant bilocals are
\begin{equation}
\begin{gathered}
   A_{kl}=\sum_{i=1}^{N}b^i(k)a_i(l), \qquad \bar A_{kl}=\sum_{i=1}^{N}a^{\dagger i}(k)b^\dagger_i(l)\ ,\\
   B_{kl}=\sum_{i=1}^{N}a^{\dagger i}(k)a_i(l), \qquad C_{kl}=\sum_{i=1}^{N}b^\dagger_i(l)b^i(k)\ .
\end{gathered}
\end{equation}
The nonzero commutators are
\begin{gather}[A_{kl},\bar A_{mn}]=N\delta_{kn}\delta_{ml}+B_{ml}\delta_{kn}+C_{kn}\delta_{ml}\ ,\\
[A_{kl},B_{mn}]=A_{kn}\delta_{ml}\ ,\\
[A_{kl},C_{mn}]=A_{ml}\delta_{kn}\ ,\\
[B_{kl},\bar A_{mn}]=\bar A_{kn}\delta_{ml},\\ [C_{kl},\bar A_{mn}] =\bar A_{ml}\delta_{kn}\ ,\\
[B_{kl},B_{mn}]=B_{kn}\delta_{ml}-B_{ml}\delta_{kn}\ ,\\
[C_{kl},C_{mn}]=C_{ml}\delta_{kn}-C_{kn}\delta_{ml}\ .
\end{gather}
And,
\begin{equation}
        [A_{kl},A_{mn}] = [\bar A_{kl},\bar A_{mn}] = [B_{kl},C_{mn}] =0 .
\end{equation}
The finite-\(N\) quadratic relations are
\begin{gather}
        \bar A A+(K-N)B-B^2=0\ , \label{app:UN-boson-constraint1} \\
        A\bar A-(C+N{\mathbb I})(C+K{\mathbb I})=0\ . \label{app:UN-boson-constraint2}
\end{gather}

\subsubsection{\texorpdfstring{$U(N)$}{U(N)} fermions}

The fermionic oscillators obey
\begin{equation}
  \{b_i(k),b^{\dagger j}(l)\}=\delta_i{}^j\delta_{kl},\qquad  \{d^i(k),d^\dagger_j(l)\}=\delta^i{}_j\delta_{kl}\ ,
\end{equation}
with all other anticommutators vanishing.  The $U(N)$-invariant bilocals are
\begin{equation}
\begin{gathered}
        A_{kl}=\sum_{i=1}^{N}d^i(k)b_i(l),  \qquad \bar A_{kl}=\sum_{i=1}^{N}b^{\dagger i}(k)d^\dagger_i(l)\ ,  \\
        B_{kl}=\sum_{i=1}^{N}b^{\dagger i}(k)b_i(l)\ ,  \qquad  C_{kl}=\sum_{i=1}^{N}d^\dagger_i(l)d^i(k)\ .
\end{gathered}
\end{equation}
The nonzero commutators are
\begin{gather}
[A_{kl},\bar A_{mn}]=N\delta_{kn}\delta_{ml}-B_{ml}\delta_{kn}-C_{kn}\delta_{ml}\ ,\\
[A_{kl},B_{mn}]=A_{kn}\delta_{ml}\ ,\\ 
[A_{kl},C_{mn}]=A_{ml}\delta_{kn}\ ,\\
[B_{kl},\bar A_{mn}]=\bar A_{kn}\delta_{ml}\ ,\\ [C_{kl},\bar A_{mn}]=\bar A_{ml}\delta_{kn}\ ,\\
[B_{kl},B_{mn}]=B_{kn}\delta_{ml}-B_{ml}\delta_{kn}\ ,\\
[C_{kl},C_{mn}]=C_{ml}\delta_{kn}-C_{kn}\delta_{ml}\ .
\end{gather}
Also,
\begin{equation}
        [A_{kl},A_{mn}] = [\bar A_{kl},\bar A_{mn}] = [B_{kl},C_{mn}] =0\ .
\end{equation}
The finite-$N$ quadratic relations are
\begin{gather}
        \bar A A-(K+N)B+B^2=0\ ,\label{app:UN-fermion-constraint1}\\
        A\bar A+(C-N{\mathbb I})(C+K{\mathbb I})=0\ . \label{app:UN-fermion-constraint2}
\end{gather}

\subsection{Summary}

The real vector models illustrate a simple rule.  The type of bilocal algebra is determined by the symmetry of the invariant pair operator in the mode labels: symmetric pairs give symplectic-type algebras, while antisymmetric pairs give orthogonal-type algebras.  For $O(N)$ models this symmetry is determined directly by the statistics of the oscillator.  For $Sp(N)$ models it is shifted by the antisymmetry of the invariant tensor $\Omega_{ij}$.  Thus $O(N)$ bosons and $Sp(N)$ fermions give symplectic-type bilocal algebras, whereas $O(N)$ fermions and $Sp(N)$ bosons give orthogonal-type bilocal algebras.

The finite-$N$ quadratic relations listed above are the operator identities that distinguish the microscopic singlet representation from the freely generated bilocal algebra.  In the main text these identities are interpreted as the defining relations of the finite-$N$ bilocal module selected by the singlet constraint.

\section{Collective field derivations and ordering conventions}
\label{app:collective}

In this appendix we explain how the finite-$N$ quadratic relations follow from the collective-field representation of the bilocal operators and explain how the Gelfand invariants are evaluated.  The computation is greatly simplified in the collective field theory representation. In this appendix repeated mode indices are summed over $1,\ldots,K$. The collective variables are invariant bilinears, and their conjugate momenta are defined with the convention
\begin{equation}
        [\alpha,\Pi]=-1 \ ,
\end{equation}
with the appropriate projection onto the symmetric, antisymmetric, or rectangular subspace. Here we are using a natural matrix notation in which the mode indices are suppressed. For the real $O(N)$ and $Sp(N)$ cases, one must distinguish between Weyl-ordered ($B^{\rm W}$) and normal-ordered ($B$) number operators. Recall that for bosons and fermions we have, respectively
\begin{equation}
        B^{\rm W}=B+{\frac N2}\mathbb I, \qquad B^{\rm W}=B-{\frac N2}\mathbb I\ . \label{app:weyl-normal-shift}
\end{equation}
In this appendix we use the Bargmann representation: creation operators are
represented by multiplication variables and annihilation operators by derivatives.
Thus the variables \(a_i,b_i\) appearing in the collective coordinates below should
be understood as Bargmann coordinates, not as the annihilation operators of
appendix~\ref{app:operator-algebras}.

\subsection{\texorpdfstring{$O(N)$}{O(N)} bosons}

For $O(N)$ bosons the collective field is given by the symmetric bilocal
\begin{equation}
        \alpha(k,l)=\sum_{i=1}^{N}a_i(k)a_i(l), \qquad \alpha(k,l)=\alpha(l,k)\ .
\end{equation}
We denote its conjugate momentum by $\Pi(k,l)$.  In the symmetric matrix notation, the basic collective representation is
\begin{equation}
        \bar A=\alpha, \qquad B=2\alpha\Pi, \qquad A=2(N-K-1)\Pi+4\Pi\alpha\Pi\ .
        \label{app:ON-boson-collective}
\end{equation}
The shift by $(K+1)$ in $A$ is the usual Jacobian contribution which arises because $\alpha$ is a symmetric $K\times K$ bilocal variable. Using \cref{app:ON-boson-collective}, one immediately obtains
\begin{equation}
        \bar A A-B^2+(1+K-N)B=0\ .
        \label{app:ON-boson-normal-constraint}
\end{equation}
In Weyl-ordered variables this becomes
\begin{equation}
        \bar A A+(1+K)B^{\rm W}-(B^{\rm W})^2={\frac N2}\left(K+1-{\frac N2}\right)\mathbb I\ .
        \label{app:ON-boson-weyl-constraint}
\end{equation}
Thus the collective representation reproduces the quadratic finite-$N$ identity used in the main text.

\subsection{\texorpdfstring{$O(N)$}{O(N)} fermions}

For $O(N)$ fermions the collective coordinate is the antisymmetric bilocal
\begin{equation}
        \beta(k,l)=\sum_{i=1}^{N}b_i(k)b_i(l)\ , 
\end{equation}
where clearly $\beta(k,l)=-\beta(l,k)$.
Let $\Pi(k,l)$ be the corresponding conjugate momentum, again projected onto the antisymmetric subspace.  The normal-ordered collective representation is
\begin{equation}
        \bar A=\beta, \qquad B=-\beta\Pi, \qquad A=-(N+K-1)\Pi-\Pi\beta\Pi\ .
        \label{app:ON-fermion-collective}
\end{equation}
The shift $(K-1)$ reflects the fact that the collective coordinate is antisymmetric. It follows from \cref{app:ON-fermion-collective} that
\begin{equation}
        \bar A A-(N+K-1)B+B^2=0\ .\label{app:ON-fermion-normal-constraint}
\end{equation}
In Weyl-ordered variables this is equivalent to
\begin{equation}
        \bar A A+(1-K)B^{\rm W}+(B^{\rm W})^2={\frac N2}\left(K-1+{\frac N2}\right) \mathbb I\ .
        \label{app:ON-fermion-weyl-constraint}
\end{equation}
This gives the collective-field derivation of the $O(N)$ fermion quadratic relation. This calculation also gives the quadratic Casimir identity.  In the $\mathfrak{so}(2K)$ oscillator realization one obtains
\begin{equation}
        {\frac 12}\Tr\left(\bar A A+A\bar A\right)+\Tr\left((B^{\rm W})^2\right)={\frac{NK}{4}}\left(N+2K-2\right)\ .
        \label{app:ON-fermion-C2}
\end{equation}
This is the invariant form of the finite-$N$ restriction.

\subsection{\texorpdfstring{$U(N)$}{U(N)} bosons}

For $U(N)$ bosons the collective coordinate is rectangular rather than symmetric or antisymmetric
\begin{equation}
        \alpha(k,l)=\sum_{i=1}^{N}a^i(k)b_i(l), \qquad \Pi(k,l)={\frac{\partial}{\partial\alpha(k,l)}}\ .
\end{equation}
The collective representation of the bilocal generators is
\begin{equation}
\begin{gathered}
        \bar A_{kl}=\alpha(k,l), \qquad B_{kl}=\alpha(k,m)\Pi(l,m)\ ,\\
        C_{kl}=\Pi(m,k)\alpha(m,l)-K\delta_{kl}\ ,  \\
        A_{kl}=(N-K)\Pi(l,k)+\Pi(m,k)\alpha(m,n)\Pi(l,n)\ .
\end{gathered}\label{app:UN-boson-collective}
\end{equation}
Substitution of \cref{app:UN-boson-collective} gives the two quadratic identities
\begin{equation}
        \bar A A+(K-N)B-B^2=0\ ,\label{app:UN-boson-first}
\end{equation}
and
\begin{equation}
        A\bar A-(C+N\mathbb I)(C+K\mathbb I)=0\ .  \label{app:UN-boson-second}
\end{equation}
These are the finite-$N$ relations for the bosonic $U(N)$ rectangular bilocal representation.

\subsection{\texorpdfstring{$U(N)$}{U(N)} fermions}

For $U(N)$ fermions we use the collective variable
\begin{equation}
        \beta(k,l)=\sum_{i=1}^{N}b^i(k)d_i(l), \qquad \Pi(k,l)={\frac{\partial}{\partial\beta(k,l)}}\ .
\end{equation}
The normal-ordered collective representation is
\begin{equation}
\begin{gathered}
        \bar A_{kl}=\beta(k,l),  \qquad B_{kl}=\beta(k,m)\Pi(l,m)\ ,  \\
        D_{kl}\equiv \sum_{i=1}^N d^i(k)d^\dagger_i(l) = N\delta_{kl} - C_{kl}= (N+K)\delta_{kl}-\Pi(m,k)\beta(m,l) \ ,  \\
        A_{kl}=(N+K)\Pi(l,k)-\Pi(m,k)\beta(m,n)\Pi(l,n)\ .
\end{gathered}\label{app:UN-fermion-collective}
\end{equation}
Here $D$ is the particle-hole conjugate number operator for the second fermion species.  In terms of the $C$ used in the main text, this is an equivalent description of the same $U(N)$ bilocal algebra. \Cref{app:UN-fermion-collective} gives
\begin{equation}
        \bar A A-(N+K)B+B^2=0\ , \label{app:UN-fermion-first}
\end{equation}
and
\begin{equation}
        A\bar A-(N+K)D+D^2=0\ .  \label{app:UN-fermion-second-D}
\end{equation}
Equivalently, in terms of \(C\),
\begin{equation}
        A\bar A+(C-N\mathbb I)(C+K\mathbb I)=0\ . \label{app:UN-fermion-second-C}
\end{equation}
The $U(N)$ fermion case also illustrates why the character description of the finite-$N$ Hilbert space is natural.  Combining the two fermion species into a single $2K$-component multiplet gives bilinears
\begin{equation}
        L_{PQ}=\bar f^{\,i}(P)f_i(Q)\ , 
\end{equation}
where $P,Q = 1,\ldots, 2K$. They obey the matrix-unit algebra
\begin{equation}
        [L_{PQ},L_{RS}] = \delta_{QR}L_{PS}-\delta_{PS}L_{RQ}\ . \label{app:UN-matrix-unit}
\end{equation}
Thus the singlet sector carries a finite-dimensional $U(2K)$-type representation. For the rectangular module relevant here one finds
\begin{equation}
        \Tr\left(L^n\right)=NK(N+K)^{n-1} \ . \label{app:UN-trace-power}
\end{equation}
The traceless part
\begin{equation}
        \widetilde L=L-{\frac{\Tr\left(L\right)}{2K}}\mathbb I\ ,
\end{equation}
has, in particular,
\begin{equation}
        \Tr\,\bigl(\widetilde L^2\bigr)={\frac12}NK(N+2K)\ , \label{app:UN-traceless-Casimir2}
\end{equation}
and 
\begin{equation}
    \Tr\,\bigl(\widetilde L^4\bigr) =
        {\frac18}NK(N+2K)\left(N^2+2KN+4K^2\right)\ .\label{app:UN-traceless-Casimir4}
\end{equation}
And in general we have, 
\begin{equation}
    \Tr\bigl(\tilde L^n\bigr) = \frac{NK(N+2K)}{2(N+K)}\sets*{\pts*{\frac{N+2K}{2}}^{n-1} - \pts*{-\frac{N}{2}}^{n-1}}\ .
\end{equation}
These formulae agree with the standard Casimir eigenvalues of the corresponding rectangular representation \cite{PismaZhETF.1.15}.

\subsection{Summary}

The collective-field calculation gives a uniform derivation of the finite-$N$ quadratic relations.  The invariant bilocals may be represented as differential operators on the collective variables, and the finite-\(N\) identities follow from the resulting non-free action of these differential operators.  In this form the origin of the shifts by $(K+1)$, $(K-1)$, and $K$ is transparent: they are the Jacobian and ordering corrections associated with symmetric, antisymmetric, and
rectangular collective coordinates respectively.  These are precisely the corrections that make the collective representation agree with the finite-$N$ bilocal representations described in the main text.

\section{Character and determinant formulae} \label{app:characters}

In this appendix we collect the character identities used in \cref{sec:characters}.  The main point is to explain how the determinant formula for the $U(N)$ fermion partition function is converted into the character of a rectangular $U(2K)$ representation.  We also record the corresponding orthogonal character and the symplectic character formulae for the $O(N)$ and $Sp(N)$ fermion models respectively.

\subsection{\texorpdfstring{$U(N)$}{U(N)} fermions: from a determinant to a rectangular character}

Let $x=e^{-\beta\omega}$. The finite-$N$ partition function of the $U(N)$ fermion model is the character of
the rectangular representation
\begin{equation}
        \lambda=(N^K)=(\underbrace{N,\ldots,N}_{K\ {\rm times}})\ ,
\end{equation}
evaluated on $K$ variables equal to $x$ and $K$ variables equal to $x^{-1}$
\begin{equation}
        Z^{\rm f}_{U(N)}(x)=s_{(N^K)}(\underbrace{x,\ldots,x}_{K},\underbrace{x^{-1},\ldots,x^{-1}}_{K})\ .
        \label{app:UN-character}
\end{equation}
Formulating the partition function as a matrix-integral, we obtain the determinant~\cite{deMelloKoch:2026dfo}
\begin{equation}
        D_{K,N}(x)=\det\left[{\frac1{i+j+1}}\,{}_2F_1\left(-N,i+j+1;i+j+2;1-x^2\right)\right]_{i,j=0}^{K-1}\ .
        \label{app:DKN-definition}
\end{equation}
The relation between these two expressions is
\begin{equation}
        D_{K,N}(x)={\frac{\prod_{m=1}^{K-1}(m!)^2}{\prod_{a,b=1}^{K}(N+a+b-1)}}\, x^{KN} s_{(N^K)}
        (\underbrace{x,\ldots,x}_{K},\underbrace{x^{-1},\ldots,x^{-1}}_{K})\ .\label{app:DKN-character-relation}
\end{equation}
Equivalently,
\begin{equation}
        Z^{\rm f}_{U(N)}(x)=\frac{\prod_{a,b=1}^{K}(N+a+b-1)}{\prod_{m=1}^{K-1}(m!)^2}\, x^{-KN}D_{K,N}(x)\ .
        \label{app:UN-partition-from-D}
\end{equation}
Thus the normalization needed to convert the determinant into the character is
\begin{equation}
 {\cal N}_{K,N}=\frac{\prod_{a,b=1}^{K}(N+a+b-1)}{\prod_{m=1}^{K-1}(m!)^2}\ . \label{app:NKN-product}
\end{equation}
Using the Barnes $G$-function, this may also be written as
\begin{equation}
        {\cal N}_{K,N} = \frac{G(N+2K+1)G(N+1)}{ G(K+1)^2G(N+K+1)^2}\ .\label{app:NKN-Barnes}
\end{equation}
We now give the proof of \cref{app:DKN-character-relation}.  Define the moments
\begin{equation}
        \mu_n =  \int_0^1 t^n\left(1-(1-x^2)t\right)^Ndt\ .\label{app:moments}
\end{equation}
Then
\begin{equation}
        \mu_n = {\frac1 {n+1}}{}_2F_1 \left( -N,n+1;n+2;1-x^2\right)\ ,
\end{equation}
and therefore
\begin{equation}
        D_{K,N}(x)=\det[\mu_{i+j}]_{i,j=0}^{K-1}\ . \label{app:DKN-Hankel}
\end{equation}
Applying Andr\'eief's identity~\cite{Forrester} gives the Hankel determinant as a Vandermonde
integral,
\begin{equation}
        D_{K,N}(x)=\frac{1}{K!}\int_{[0,1]^K} \Delta(t)^2 \prod_{\alpha=1}^{K}\left(1-(1-x^2)t_\alpha\right)^N
        dt_1\cdots dt_K\ ,\label{app:DKN-Vandermonde}
\end{equation}
where
\begin{equation}
        \Delta(t)=\prod_{\alpha<\beta}(t_\beta-t_\alpha)\ .
\end{equation}
Now make the change of variables
\begin{equation}
        z_\alpha=x^{-1}+(x-x^{-1})t_\alpha\  . \label{app:z-change}
\end{equation}
Then it follows
\begin{equation}
        1-(1-x^2)t_\alpha=xz_\alpha, \qquad dt_\alpha=\frac{dz_\alpha}{x-x^{-1}}, \qquad
        \Delta(t)=\frac{\Delta(z)}{(x-x^{-1})^{K(K-1)/2}}\ .
\end{equation}
Hence
\begin{equation}
        D_{K,N}(x) = \frac{x^{KN}}{(x-x^{-1})^{K^2}} H_{K,N}(x,x^{-1})\ , \label{app:DKN-HKN}
\end{equation}
where
\begin{equation}
        H_{K,N}(u,v) = \det\left[ \frac{u^{N+a+b-1}-v^{N+a+b-1}}{N+a+b-1}\right]_{a,b=1}^{K}\ .
        \label{app:HKN-definition}
\end{equation}
All that remains is to relate \(H_{K,N}(u,v)\) to a Schur polynomial.  The required identity is
\begin{equation}
        H_{K,N}(u,v) = \frac{\prod_{m=1}^{K-1}(m!)^2}{\prod_{a,b=1}^{K}(N+a+b-1)} (u-v)^{K^2} s_{(N^K)}
        (\underbrace{u,\ldots,u}_{K},\underbrace{v,\ldots,v}_{K})\ . \label{app:HKN-Schur}
\end{equation}
This follows from the Weyl character formula by taking the confluent limit in which the first $K$ variables approach $u$ and the remaining $K$ variables approach $v$.  The factor $(u-v)^{K^2}$ is the part of the $2K$-variable Vandermonde which mixes the two groups of variables, while the remaining numerical factor comes from the two confluent Vandermonde determinants and from the exponents associated with the rectangular diagram $(N^K)$. Combining \cref{app:DKN-HKN,app:HKN-Schur}, and setting $u=x$, $v=x^{-1}$, gives \cref{app:DKN-character-relation}.  This proves that the determinant expression is exactly the rectangular Schur character, up to the normalization in \cref{app:UN-partition-from-D}.

\subsection{\texorpdfstring{$O(N)$}{O(N)} fermions: the orthogonal character}

The same representation-theoretic logic gives a compact expression for the $O(N)$ fermion partition function.  The Cartan generators of the $\mathfrak{so}(2K)$ bilocal algebra may be chosen as
\begin{equation}
        H_k = {\frac N2} - \sum_{i=1}^{N}b_i^\dagger(k)b_i(k)\ , \label{app:ON-Cartan}
\end{equation}
Thus the Fock vacuum has highest weight
\begin{equation}
        \lambda = \left({\frac N2},{\frac N2},\ldots,{\frac N2}\right)\ . \label{app:ON-highest-weight}
\end{equation}
For the Hamiltonian
\begin{equation}
        H=\omega\sum_{k=1}^{K}H_k\ ,
\end{equation}
the partition function is therefore the corresponding even orthogonal character:
\begin{equation}
        Z^{\rm f}_{O(N)}(x) = x^{NK/2} o^{+}_{\lambda} ( \underbrace{x^{-1},\ldots,x^{-1}}_{K} )\ , \label{app:ON-character-partition}
\end{equation}
with $x=e^{-\beta\omega}$. Here $o^{+}_{\lambda}$ denotes the $SO(2K)$ character with positive chirality (as determined by the sign of the weight component $\lambda_K$).
In determinant form,
\begin{equation}
        o^{+}_{\lambda}(x_1,\ldots,x_K)= \frac{ \det\left[ x_j^{\lambda_i+K-i} + x_j^{-(\lambda_i+K-i)} \right]
        + \det\left[ x_j^{\lambda_i+K-i} - x_j^{-(\lambda_i+K-i)} \right]}{\det\left[ x_j^{K-i} + x_j^{-(K-i)}\right]}\ ,
        \label{app:orthogonal-character}
\end{equation}
with $i,j=1,\ldots,K$.  This formula is to be understood by the usual limiting procedure when all variables are set equal. At $x=1$, the partition function reduces to the dimension of the $\mathfrak{so}(2K)$ representation with highest weight $\lambda$.  Weyl's dimension formula gives
\begin{equation}
  \dim\left({\frac N2},{\frac N2},\ldots,{\frac N2}\right) = \prod_{1\leq i<j\leq K}\frac{N+2K-i-j}{2K-i-j}\ .
        \label{app:ON-dimension}
\end{equation}
This agrees with the finite-$N$ state counting obtained directly from the bilocal creation operators.

\subsection{\texorpdfstring{$Sp(N)$}{Sp(N)} fermions: the symplectic character}

We choose the Cartan generators of the $\mathfrak{sp}(2K)$ bilocal algebra of the form,
\begin{equation}
    H_k = \frac12\pts*{L_{kk} + L_{(K+k)(K+k)}}\ ,\label{app:SpN-Cartan}
\end{equation}
Thus the Fock vacuum now has the lowest weight, 
\begin{equation}
    \lambda = \pts*{-\frac{N}{2},-\frac{N}{2},\dots, -\frac{N}{2}}\ .
\end{equation}
Recall that for the even symplectic group, the lowest weight state is exactly the negative of the highest weight state \emph{i.e.} for the highest weight state we have, 
\begin{equation}
    \lambda = \pts*{\frac{N}{2},\frac{N}{2},\dots, \frac{N}{2}}\ .
\end{equation}
For the Hamiltonian,
\begin{equation}
    H = \omega\sum_{k=1}^K H_k\ ,
\end{equation}
the partition function is the corresponding symplectic character:
\begin{equation}
    Z^{\text{f}}_{Sp(N)}(x) = x^{-NK/2}{sp}_\lambda(\underbrace{x^{-1},\dots,x^{-1}}_K)\;\;,\qquad x = e^{-\beta\omega}\ , 
\end{equation}
where $sp_\lambda$ is the $Sp(2K)$ character. A simple check is to count the number of independent states generated by $\bar A$'s at fixed particle number.  The results for small values of $N$ and $K$ are shown in \cref{tab:SpAdim}. 
\begin{table}[H]
\centering
\begin{tabular}{c|ccccccc}
\hline
\((\frac{N}{2},K)\) & \((1,2)\) & \((1,3)\) & \((1,4)\) & \((1,5)\) &
\((2,2)\) & \((2,3)\) & \((3,2)\)\\
\hline
\(n_P=0\)  & 1 & 1 & 1  & 1  & 1 & 1  & 1  \\
\(n_P=2\)  & 3 & 6 & 10 & 15 & 3 & 6  & 3  \\
\(n_P=4\)  & 1 & 6 & 20 & 50 & 6 & 21 & 6  \\
\(n_P=6\)  &   & 1 & 10 & 50 & 3 & 28 & 10 \\
\(n_P=8\)  &   &   & 1  & 15 & 1 & 21 & 6  \\
\(n_P=10\) &   &   &    & 1  &   & 6  & 3  \\
\(n_P=12\) &   &   &    &    &   & 1  & 1  \\
\hline
\end{tabular}
\caption{Number of the independent states generated by the Sp\((N)\)
fermion bilocal creation operators \(\bar A\) at fixed particle number
\(n_P\). As earlier, blank entries denote levels which are not present for the corresponding values of $N$ and $K$.}
\label{tab:SpAdim}
\end{table}
In determinant form,
\begin{equation}
    sp_\lambda(x_1,\dots, x_K) = \frac{\det\sq*{x_j^{\lambda_i + K - i + 1} - x_j^{-(\lambda_i + K - i + 1)}}^K_{i,j=1}}{\det\sq*{x_j^{K-i+1} - x_j^{-(K-i+1)}}^K_{i,j=1}}\ .
\end{equation}
Note that to evaluate this formula, we must substitute the \emph{highest} weight $\lambda$ rather than the weight of the vacuum unlike earlier. At $x=1$, the partition function reduces to the dimension of the $\mathfrak{sp}(2K)$ representation with highest weight $\lambda$. From Weyl's dimension formula once again, we have
\begin{equation}
    \dim\pts*{\frac N2, \frac N2,\dots, \frac N2} = \left(\prod_{i=1}^K\frac{\frac N2 + K - i + 1}{K - i + 1}\right) \left(\prod_{1\leq i < j \leq K}\frac{N + 2K - i - j + 2}{2K - i - j + 2}\right)\ .
\end{equation}
As before this agrees with the finite-$N$ state counting obtained directly in ref.~\cite{deMelloKoch:2026dfo}.

\end{document}